\documentclass[prl,aps,twocolumn,superscriptaddress,preprintnumbers,showpacs,floatfix,amsmath,amssymb]{revtex4}

\usepackage{graphicx}
\usepackage{dcolumn}
\usepackage{bm}
\usepackage{amsmath}
\usepackage{graphicx}
\usepackage{amssymb}
\usepackage{mathrsfs}
\usepackage{bbm}
\usepackage{epsfig}

\newcommand{\be}{\begin{eqnarray}}
\newcommand{\ee}{\end{eqnarray}}

\begin{document}

%
%
%
\title{ Solitons and Collapse in the  $\lambda$-repressor protein    }

\author{Andrey Krokhotin}
\email{Andrei.Krokhotine@cern.ch}
\affiliation{Department of Physics and Astronomy, Uppsala University,
P.O. Box 803, S-75108, Uppsala, Sweden}
\author{Martin Lundgren}
\email{martin.lundgren@gmail.com}
\affiliation{Department of Physics and Astronomy, Uppsala University,
P.O. Box 803, S-75108, Uppsala, Sweden}
\author{Antti J. Niemi}
\email{Antti.Niemi@physics.uu.se}
\affiliation{Department of Physics and Astronomy, Uppsala University,
P.O. Box 803, S-75108, Uppsala, Sweden}
\affiliation{
Laboratoire de Mathematiques et Physique Theorique
CNRS UMR 6083, F\'ed\'eration Denis Poisson, Universit\'e de Tours,
Parc de Grandmont, F37200, Tours, France}

\begin{abstract}
\noindent
The  enterobacteria lambda phage is a paradigm temperate bacteriophage. 
Its lysogenic and lytic life cycles echo competition between the
DNA binding $\lambda$-repressor (CI) and CRO proteins. Here we  scrutinize the structure, 
stability and folding pathways of the $\lambda$-repressor protein, that controls the transition 
from the lysogenic  to the lytic state. We first investigate the
super-secondary helix-loop-helix composition of its
backbone. We use a discrete Frenet framing to 
resolve the backbone spectrum in terms of bond and torsion angles.
Instead of four, 
there  appears to be seven individual loops. We model the putative loops 
using an explicit soliton Ansatz. It is based on
the standard soliton profile of the  continuum nonlinear Schr\"odinger equation.
The accuracy of the Ansatz far exceeds the B-factor fluctuation distance 
accuracy of the experimentally  determined protein configuration. 
We then investigate  the folding pathways  and dynamics 
of the $\lambda$-repressor protein.  We introduce a coarse-grained energy function to model 
the backbone in terms of  the C$_\alpha$
atoms and the side-chains in terms of the relative orientation of the C$_\beta$ atoms.
We describe the folding dynamics in terms of relaxation dynamics, and find
that the folded configuration can be reached from a very generic initial configuration.
We conclude that folding is dominated by the temporal ordering of soliton formation. 
In particular, the third soliton should  appear before the first and second. 
Otherwise, the  DNA binding turn does not  acquire its correct structure. We confirm 
the stability of the folded configuration by repeated heating and cooling simulations. 
\end{abstract}


\maketitle

\section{I: Introduction}


The transition between the lysogenic and the lytic state in bacteriophage $\lambda$ infected {\it E. coli} cell 
is the paradigm genetic switch mechanism. It is described  in numerous molecular biology textbooks and 
review articles \cite{lambda}-\cite{lambda6}.  
The interplay between the lysogeny maintaining $\lambda$-repressor (CI) protein and  the CRO  regulator protein that 
controls the transition to the lytic state is a simple model for more complex regulatory networks, including 
those that can  lead to cancer in humans.

In the present article we describe the physical properties of the $\lambda$-repressor protein, that controls 
the lysogenic-to-lytic transition. 
We investigate in detail 
the stability of its native conformation, the dynamics 
of the folding process, and the landscape of  folding pathways.  
We find that the folded configuration displays
a structure which is unique among all known protein structures. We also conclude 
that the folding pathways are entirely dominated by the loop
regions.  In particular, the temporal ordering of loop formation appears to be the decisive 
factor for the protein's ability to reach its native fold. If 
solitons form in a wrong order the protein may misfold.


Full crystallographic information of the experimental $\lambda$-repressor structure that we use in our investigation 
is available in  Protein Data Bank (PDB) \cite{pdb} 
under the code 1LMB.   This structure is a homo-dimer with 92 residues in each of  
the two monomers. It maintains the lysogenic state by binding  to DNA with a helix-turn-helix 
motif that is  located between  the  residue sites  33-51. Throughout this article we shall 
use  the PDB  indexing
of the residues.  

For the statistical analyses that are presented here, we utilize a subset of PDB data that consists of the canonical
set of structures with better than 2.0 \AA  ~ resolution. We have compared the results 
with the subset that contains only those proteins with better than 2.0 \AA ~ resolution and
with less than 30$\%$ sequence similarity. Our conclusions are independent of the 
data set, and for illustrative purposes we here use the canonical 2.0 \AA ~ set. 

This article is composed as follows: We first explain how to describe the geometry of a generic
folded protein in terms of its backbone central C$_\alpha$ carbons.  We propose that a coarse-grained
energy function, that models the backbone geometry, can be constructed with only 
the C$_\alpha$ coordinates as dynamical variables. We argue that a variant 
of the discrete non-linear Schr\"odinger (DNLS) equation is a suitable {\it Master Equation} to describe 
folded proteins, in terms of its dark soliton solution. We then proceed to utilize this general framework
to study the structure of the $\lambda$-repressor protein. We show that the $\lambda$-repressor backbone is composed 
from seven individual soliton solutions of the DNLS equation, within the accuracy of 
crystallographic structure measurements. In the same vein we propose, that protein 
folding can be described in terms of a coarse grained model, based on relaxation dynamics.
We utilize this to investigate the folding dynamics of the $\lambda$-repressor.
We conclude that the temporal ordering of soliton formation is important for reaching the
correct native state. A wrong ordering in soliton formation can be a cause for misfolding.
We observe that the second soliton has a peculiar structure that sets it apart from any other
known structure in all proteins. 
%
%
%
%
%
%
%
%
%
%
%
%
%
%
%

\section{II: Methods:}

\subsection{A: Backbone geometry}

Our  analysis  of the $\lambda$-repressor protein  will be 
based on an effective, coarse grained  energy function approach that has been recently 
developed  in \cite{ulf}-\cite{hu}. In this approach the protein geometry 
is described in terms of the backbone C$_\alpha$ atoms. The ensuing 
bond and torsion angles assume  the r\^ole 
of the dynamical variables. 
These  angles are constructed as follows:
Let $\mathbf r_i$ be the coordinate sites of the C$_\alpha$ carbons, where the 
index $i=1,...,N$ runs over all amino acids. 
For a given protein these coordinates can be read from the PDB. 
For each site $i$, we introduce the unit tangent vector 
\begin{equation}
\mathbf t_i = \frac{ {\bf r}_{i+1} - {\bf r}_i  }{ |  {\bf r}_{i+1} - {\bf r}_i | }
\label{t}
\end{equation}
the unit binormal vector
\begin{equation}
\mathbf b_i = \frac{ {\mathbf t}_{i-1} - {\mathbf t}_i  }{  |  {\mathbf t}_{i-1} - {\mathbf t}_i  | }
\label{b}
\end{equation}
and the unit normal vector 
\begin{equation}
\mathbf n_i = \mathbf b_i \times \mathbf t_i
\label{n}
\end{equation}
The orthogonal triplet ($\mathbf n_i, \mathbf b_i , \mathbf t_i$) determines a discrete version of the Frenet  
frame at each position $\mathbf r_i$ of the backbone. 

The backbone bond angles are
\begin{equation}
\kappa_{i} \ \equiv \ \kappa_{i+1 , i} \ = \ \arccos \left( {\bf t}_{i+1} \cdot {\bf t}_i \right)
\label{bond}
\end{equation}
and the backbone torsion angles are
\begin{equation}
\tau_{i} \ \equiv \ \tau_{i+1,i} \ = \ {\rm sign}\{ \mathbf b_{i-1} \times \mathbf b_i \cdot \mathbf t_i \}
\cdot \arccos\left(  {\bf b}_{i+1} \cdot {\bf b}_i \right) 
\label{tors}
\end{equation}
Conversely, if these angles are all known, we can use 
\begin{equation}
\left( \begin{matrix} {\bf n}_{i+1} \\  {\bf b }_{i+1} \\ {\bf t}_{i+1} \end{matrix} \right)
= 
\left( \begin{matrix} \cos\kappa \cos \tau & \cos\kappa \sin\tau & -\sin\kappa \\
-\sin\tau & \cos\tau & 0 \\
\sin\kappa \cos\tau & \sin\kappa \sin\tau & \cos\kappa \end{matrix}\right)_{\hskip -0.1cm i+1 , i}
\left( \begin{matrix} {\bf n}_{i} \\  {\bf b }_{i} \\ {\bf t}_{i} \end{matrix} \right) 
\label{DFE2}
\end{equation}
to iteratively construct the frame at position $i+i$ 
from the frame at position $i$. Once we have all the frames, we obtain the entire backbone from
\begin{equation}
\mathbf r_k = \sum_{i=0}^{k-1} |\mathbf r_{i+1} - \mathbf r_i | \cdot \mathbf t_i
\label{dffe}
\end{equation}
Without any loss of generality we may set $\mathbf r_0 = 0$, and choose $\mathbf t_0$ to 
point into the direction of the positive $z$-axis.

We note that the relation (\ref{dffe}) does not involve the vectors $\mathbf n_i$ and $\mathbf b_i$. 
Consequently we may rotate the ($\mathbf n_i, \mathbf b_i$) frame vectors, without affecting 
the backbone, by selecting an arbitrary 
linear combination of these two vectors independently at each site $i$. 
For this we introduce a local SO(2) transformation that rotates the 
($\mathbf n_i, \mathbf b_i$)  by an angle 
$\Delta_i$ so that the $\mathbf t_i$ remain intact,
\begin{equation}
 \left( \begin{matrix}
{\bf n} \\ {\bf b} \\ {\bf t} \end{matrix} \right)_{\!i} \!
\rightarrow  \!  e^{\Delta_i T^3} \left( \begin{matrix}
{\bf n} \\ {\bf b} \\ {\bf t} \end{matrix} \right)_{\! i} =   \left( \begin{matrix}
\cos \Delta_i & \sin \Delta_i & 0 \\
- \sin \Delta_i & \cos \Delta_i & 0 \\ 
0 & 0 & 1  \end{matrix} \right) \left( \begin{matrix}
{\bf n} \\ {\bf b} \\ {\bf t} \end{matrix} \right)_{\! i}
\label{discso2}
\end{equation}
where the  SO(3) generators are $(T^i)_{jk} = \epsilon_{ijk} $,
\[
[T^i , T^j ] = \epsilon_{ijk}T^k
\]
We combine $\mathbf n$ and $\mathbf b$ into the complex vector
\[
\mathbf n + i \mathbf b
\]
and rewrite (\ref{discso2}) as
\begin{equation}
\mathbf n_i + i \mathbf b_i \to e^{i \Delta_i} (\mathbf n_i + i \mathbf b_i ) \equiv \mathbf e^1_i + i \mathbf e^2_i
\label{e12}
\end{equation}
The frame rotation (\ref{discso2}) corresponds to the following transformation
in the bond and torsion angles, 
\begin{equation}
\kappa_{i}  \ T^2  \ \to \  e^{\Delta_{i} T^3} ( \kappa_{i} T^2 )\,  e^{-\Delta_{i} T^3}
\label{sok}
\end{equation}
\begin{equation}
\tau_{i}  \ \to \ \tau_{i} + \Delta_{i-1} - \Delta_{i}
\label{sot}
\end{equation}
Since the transformation (\ref{sok}), (\ref{sot}) leaves $\mathbf t_i$ intact 
it has no  effect on the backbone. 

{\it A priori}, the fundamental range of the bond angle $\kappa_i$ is  $\kappa_i \in [0,\pi]$. For the 
torsion angle the range is $\tau_i \in [-\pi, \pi)$. Consequently we may 
identify ($\kappa_i, \tau_i$) with the canonical 
latitude and longitude angles of a two-sphere $\mathbb S^2$. 
However, in the sequel we find it useful to extend the range
of $\kappa_i$ into $ [-\pi,\pi]$ $mod(2\pi)$, but with no change in the range of $\tau_i$. 
We compensate for this two-fold covering of $\mathbb S^2$ 
by introducing the following discrete $\mathbb Z_2$ symmetry
\begin{equation}
\begin{matrix}
\ \ \ \ \ \ \ \ \ \kappa_{k} & \to  &  - \ \kappa_{k} \ \ \ \hskip 1.0cm  {\rm for \ \ all} \ \  k \geq i \\
\ \ \ \ \ \ \ \ \ \tau_{i }  & \to &  \hskip -2.5cm \tau_{i} - \pi 
\end{matrix}
\label{dsgau}
\end{equation}
We note that this is a special case of (\ref{sok}), (\ref{sot}), with
\[
\begin{matrix} 
\Delta_{k} = \pi \hskip 1.0cm {\rm for} \ \ k \geq i+1 \\
\Delta_{k} = 0 \hskip 1.0cm {\rm for} \ \ k <  i+1 
\end{matrix}
\]

The regular protein secondary 
structures  correspond to definite values of
$(\kappa_i, \tau_i)$.  For example standard $\alpha$-helix is
\begin{equation}
\alpha-{\rm helix:} \ \ \ \ \left\{ \begin{matrix} \kappa \approx \frac{\pi}{2}  \\ \tau \approx 1\end{matrix} \right.
\label{bc1}
\end{equation}
and standard $\beta$-strand is 
\begin{equation}
\beta-{\rm strand:} \ \ \ \ \left\{ \begin{matrix} \kappa \approx 1 \\ \tau \approx \pi \end{matrix}  \right.
\label{bc2}
\end{equation}
Similarly we can describe all the other regular secondary structures such as 3/10 helices, 
left-handed helices {\it etc.} with definite constant values of $\kappa_i$ and $\tau_i$.
Loops are configurations that interpolate between these regular structures, so that 
along a loop the values of ($\kappa_i, \tau_i$) are variable.

Finally, we compute the average value of the  bond length in (\ref{dffe}) using PDB.
The result shown in Figure 1 
\begin{figure}
  \begin{center}
    \resizebox{7.5cm}{!}{\includegraphics[]{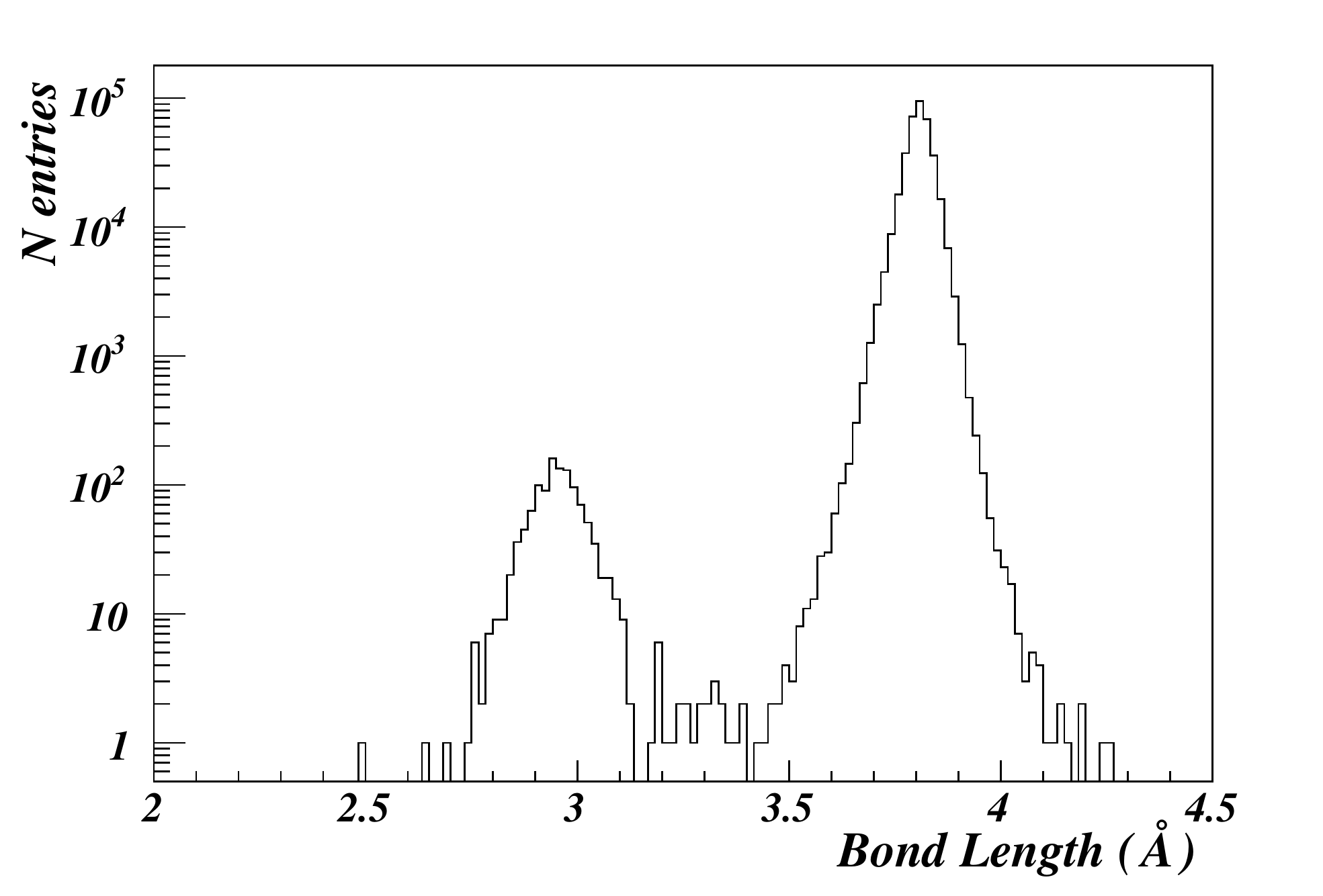}}
    \caption{The distribution of the C$_\alpha$-C$_\alpha$ bond length in PDB. The fluctuations around
    the average value $d \approx 3.8$ \AA ~ are small, the secondary peak around 2.9 \AA~ 
    is due  to {\it cis}-proline. 
    Note logarithmic scale.}
    \label{fig:simple}
  \end{center}
\end{figure}
is constructed using the canonical set of protein structures which are
measured with better than 2.0 \AA ~ resolution. The average value is concentrated around 
\begin{equation}
 |\mathbf r_{i+1} - \mathbf r_i | = d  \approx  3.8 \, {\mathrm \AA}
 \label{dist}
 \end{equation}
In our theoretical analysis we use the fixed bond length value (\ref{dist}).  We  also 
impose the forbidden volume (steric) constraint
 \begin{equation}
 |\mathbf r_{i} - \mathbf r_k | \geq  3.8 \ \dot {\mathrm A}  \ \ \ \ {\rm for} \ \ |i-k| \geq 2 
\label{fvol}
\end{equation}
between the backbone C$_\alpha$ atoms. 
This condition is well respected by folded protein structures in PDB.

%
%
%
%
%
%
%
%
%
%
%
%
%
%
%

\subsection{B: Side-chain geometry}

Following \cite{hu} we
 characterize the side-chain directions  in terms of  
directional vectors $\mathbf u_i$ that point  from the C$_\alpha$ towards the
corresponding C$_\beta$ carbon. At each C$_\alpha$
\begin{equation}
\mathbf u_i = \left( \begin{matrix} \sin\theta_i \cdot \cos \varphi_i \\ \sin\theta_i \cdot \sin \varphi_i 
\\ \cos \theta_i \end{matrix} \right) 
\label{uvec}
\end{equation}
The latitude angle $\theta_i$ counts deviation from the direction of the corresponding  Frenet frame
tangent vector $\mathbf t_i$. When $\theta_i=0$ the $\mathbf t_i$ and $\mathbf u_i$ are parallel.
Note that the angle $\theta_i$
remains  invariant under the rotation (\ref{discso2}).
We can compute the values of $\theta_i$ from the PDB. As shown in Figure 1 the
range of variations in $\theta_i$ are quite small,  it fluctuates around
\begin{equation}
<\theta> \ \approx \ 1.98 \ ({\rm rad})  
\label{thetave}
\end{equation}

The longitude $\varphi_i$ in (\ref{uvec}) measures distance
from the direction of the Frenet 
frame normal vector $\mathbf n_i$. It is the azimuthal angle between $\mathbf n_i$
and the projection of $\mathbf u_i$ on the normal plane spanned by ($\mathbf n_i, \mathbf b_i$).
Under the frame rotation (\ref{discso2}) we  have
\begin{equation}
\varphi_i \  \to \ \varphi_i + \Delta_i
\label{varp1}
\end{equation}
and consequently the values of $\varphi_i$ depend on the framing. 
From PDB we find that in the Frenet frames the values of $\varphi_i$ 
are  subject to relatively small fluctuations around the average  value
\begin{equation}
<\varphi> \ \approx \ - 2.43 \ ({\rm rad})  
\label{varp2}
\end{equation}
\begin{figure}
  \begin{center}
    \resizebox{7.5cm}{!}{\includegraphics[]{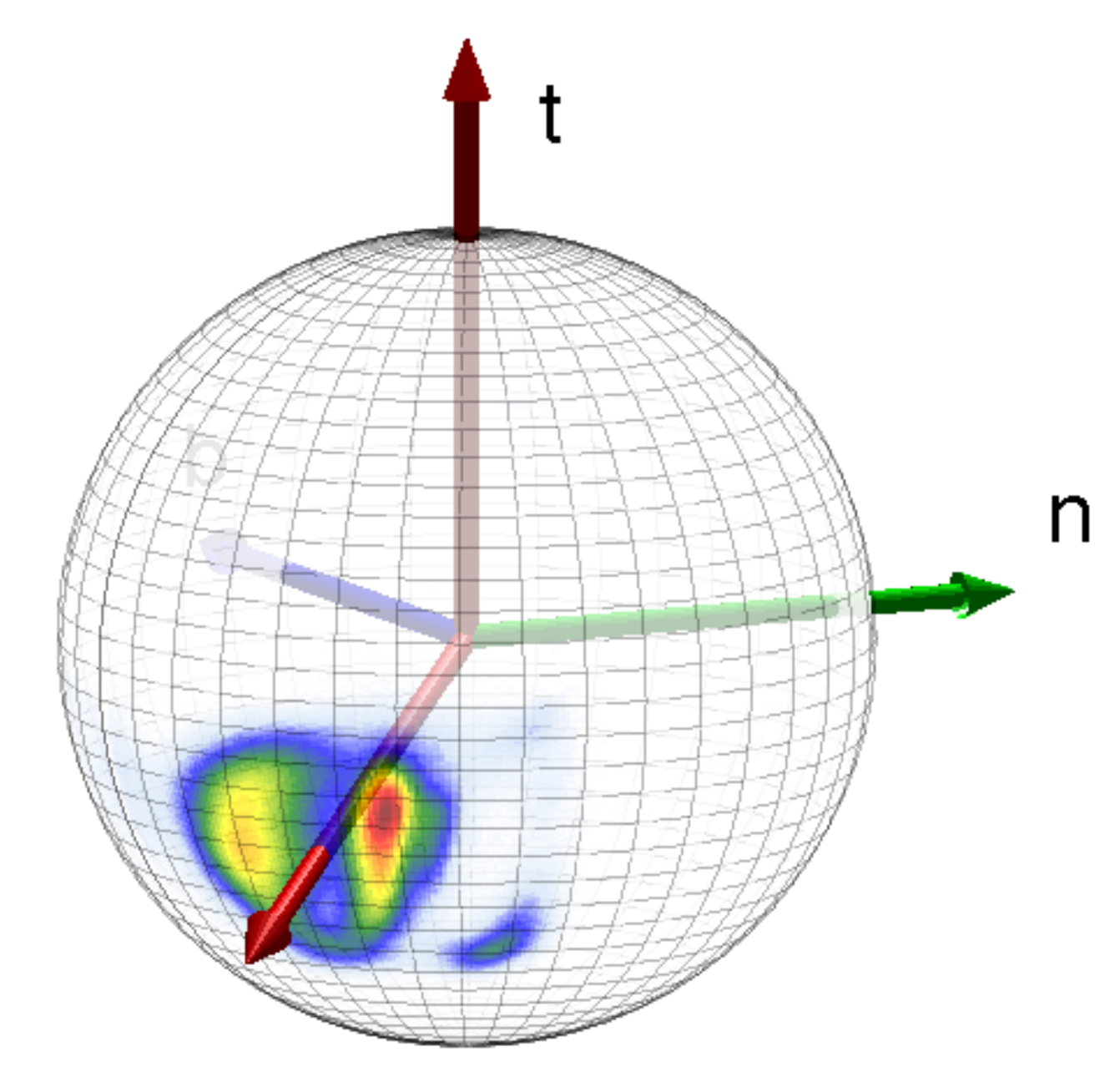}}
    \caption{(Color online)  In the Frenet frames the fluctuations in the direction 
    of the vectors $\mathbf u_i$ around their average value,
    given by (\ref{thetave}), (\ref{varp2}) and denoted by the (red) lateral vector, 
    are relatively small. The left-hand side of the horse-shoe corresponds to
    $\beta$-strands, the right-hand side to $\alpha$-helices and the connecting
    region at the bottom  corresponds to loops. The minuscule isolated island corresponds to 
    the "left-handed $\alpha$" region.  }
    \label{fig:simple}
  \end{center}
\end{figure}

As shown in Figure 2, it is remarkable that the direction of $\mathbf u_i$  nutates in a very regular horse-shoe shaped manner,
that reflects the underlying secondary structure environment.
This proposes that there is a strong local coupling between the two angular variables
$\theta_i$ and $\varphi_i$, that depends on the secondary structure.

The notable expection from (\ref{thetave}), (\ref{varp2}) is the left-handed loop region. It is visible in
Figure 2 as a minuscule isolated region, with 
\begin{equation}
\begin{matrix} 
<\theta> \ \approx \ 2.25 \ ({\rm rad}) \\
<\varphi> \ \approx \ - 1.90 \ ({\rm rad})  
\end{matrix}
\end{equation}
But this is also {\it quite} close to the values in (\ref{thetave}), (\ref{varp2}).
%
%
%
%
%
%
%
%
%
%
%
%
%
%
%

\subsection{C: Backbone  energy and solitons}

Proteins display a hierarchy which is determined by the spatial length scale. As the length scale increases, 
shorter distance dynamical variables become gradually disengaged.  Following the general
concept of universality \cite{widom}-\cite{fisher}, we utilize this hierarchy of scales 
to systematically coarse grain the energy function. At each level of hierarchy we 
retain explicitly only those variables that remain relevant. 
The variables that are  less and 
less so as the distance scale increases are accounted for effectively, through the functional 
form and  the values of parameters  in the various  
individual energy contributions that involve the remaining relevant variables only.

In  \cite{ulf}-\cite{hu}, see also \cite{oma},  it has been argued
that a Landau free energy function that aims to compute the overall fold geometry can be 
based solely on those variables that determine the positions of the central C$_\alpha$ atoms.
Since the fluctuations in the bond lengths are minimal, see Figure 1,
the leading order contribution to the energy then 
involves only the  bond and torsion angles for the C$_\alpha$  backbone. 
The functional form of the  energy
can be uniquely determined by symmetry considerations. For this we note that
{\it any} backbone energy function 
that involves only the bond and torsion angles must remain invariant under the SO(2) transformation
(\ref{sok}), (\ref{sot}).  Indeed, previously it has  already been shown \cite{ulf}-\cite{hu}  how 
this SO(2) gauge invariance allows us to uniquely deduct
the functional form of the energy function,  in the long distance limit where any higher order non-local
contribution can be ignored. In this limit  the energy function can only involve the following terms 
\cite{ulf}-\cite{nora},
\[
E = - \sum\limits_{i=1}^{N-1}  2\, \kappa_{i+1} \kappa_i  + \sum\limits_{i=1}^N
\biggl\{  2 \kappa_i^2 + q\cdot (\kappa_i^2 - m^2)^2  
\]
\begin{equation}
\left.  + \frac{d_\tau}{2} \, \kappa_i^2 \tau_i^2   -  b_\tau  \kappa_i^2  \tau_i  - a_\tau \tau_i     +  
\frac{c_\tau}{2}  \tau^2_i 
\right\} 
\label{E1}
\end{equation}
The detailed derivation of  (\ref{E1}) is presented in  \cite{ulf}. It involves the introduction
of frame (gauge) invariant combinations of the 
Frenet frame bond and torsion angles \cite{oma}, \cite{faddeev2}.
For the present purposes it suffices to observe, that (\ref{E1}) coincides with the one dimensional
discretized Abelian Higgs Model Hamiltonian in the unitary gauge, in terms of 
the Frenet frame bond and torsion variables.

We can also recognize 
(\ref{E1}) as a variant of the discrete nonlinear Schr\"odinger  (DNLS) equation \cite{faddeev}:
The first sum together with the three first terms in the second sum comprise  exactly   the 
energy of  the standard  DNLS equation  when expressed in terms of the  
Hasimoto  variable of fluid mechanics \cite{nora}, \cite{faddeev}.  The fourth ($b_\tau$) is a 
conserved quantity in the DNLS hierarchy,  the   "momentum",  and the fifth term ($a_\tau$) is the 
conserved "helicity".  
The last ($c_\tau$) term is the (non-conserved)  Proca mass term that we include for completeness.

The energy function (\ref{E1}) does not purport to explain the fine 
details of the atomary level mechanisms that give rise to protein folding. 
Rather, it examines the properties of a folded protein 
backbone in terms of universal physical arguments along the lines of  
\cite{widom}-\cite{fisher}. Indeed,
the functional form (\ref{E1}) is deeply anchored in the elegant mathematical structure of
integrable hierarchies \cite{faddeev}. Within this framework no term beyond those
in the integrable hierarchy can be added  without violating 
the underlying general and  elegant mathematical principles.
In this sense, 
(\ref{E1})  is the {\it universal }
long distance limit that would emerge from {\it any} microscopic level
Schr\"odinger operator, whenever we truncate the Landau free energy 
to explicitly include only the backbone bond and torsion angles. 

Remarkably, we have found that despite the very general nature of argumentation that
leads us to adopt the energy function (\ref{E1}), it is fully capable of describing folded protein 
backbones with sub-atomic precision of around 0.5 \AA, and even better \cite{peng2}. 
This is due to the observation \cite{maxim}, \cite{nora}, that (\ref{E1}) describes proteins in
terms of solitons that are the paradigm structural self-organizers. Indeed, solitons are
tremendously robust in 
their ability to preserve the form under both quantum mechanical and thermal fluctuations. 

We derive the relevant soliton profile as follow: We first introduce the $\tau$-equation of motion 
\[ 
\frac{\partial E}{\partial \tau_i} =  d_\tau \kappa_i^2 \tau_i - b_\tau \kappa_i^2 - a_\tau + c_\tau \tau_i = 0 
\]
from which we solve
\begin{equation}
\tau_{i} [\kappa] = \frac{a_\tau + b_\tau \kappa_i^2}{c_\tau + d_\tau \kappa^2_{i} }  
\label{tauk} 
\end{equation}
Even though there are four parameters in (\ref{tauk}) one of them, the overall scale, cancels out.
In the sequel we shall choose  $a_\tau=-1.0$ so that for an $\alpha$-helix (\ref{bc1}) we have
\[
\tau_{i} [\alpha] =  \frac{1 + b_\tau \kappa_i^2}{c_\tau + d_\tau \kappa^2_{i} }  \approx 1  \ \mod(2\pi) 
\]
and for a $\beta$-strand (\ref{bc2})
\[
\tau_i [\beta] =  \frac{1 + b_\tau \kappa_i^2}{c_\tau + d_\tau \kappa^2_{i} }  \approx \pi \  \mod(2\pi) \ \ 
\]

When we  use (\ref{tauk}) to eliminate the torsion angles we get 
for the bond angles the energy 
\begin{equation}
E[\kappa] = - \sum\limits_{i=1}^{N-1}  2\, \kappa_{i+1} \kappa_{i}  + \sum\limits_{i=1}^N
 2 \kappa_{i}^2 + V[\kappa_i]  
\label{Ekappa}
\end{equation}
where
\[
V[\kappa]  =  - \left( \frac{b_\tau c_\tau - a_\tau d_\tau}{d_\tau}  \right) \cdot \frac{1}{c_\tau+d_\tau\kappa^2} 
\]
\begin{equation}
- \left( \frac{b_\tau^2 + 8 q m^2}{2b_\tau} \right)
\cdot \kappa^2 + q\cdot \kappa^4
\label{Vkappa}
\end{equation}
The functional form of (\ref{Vkappa}) is familiar from various studies in mathematical physics: The
first term relates  to the potential energy in a Calogero-Moser system.
The second and third terms have the 
conventional form of a symmetry breaking double-well potential.  Depending on 
the parameter values we may  
be either in the broken symmetry phase where $\kappa$ and $\tau$  both acquire a 
non-vanishing and constant  ground state value, or  in the symmetric  phase 
where $\kappa $ vanishes.  

In applications to proteins,  regular structures such as 
helices  (\ref{bc1}) and strands (\ref{bc2}) correspond to different broken symmetry
ground states of the energy. Furthermore, the numerical value of the first term in (\ref{Vkappa})
is often small in comparison to the second and third, and so is $b_\tau^2$ in comparison to
$8qm^2$ so that for an $\alpha$-helix (\ref{bc1}) 
\begin{equation}
m \approx \frac{\pi}{2}
\label{malpha}
\end{equation}
and for a $\beta$-strand
\begin{equation}
m \approx 1.0
\label{mbeta}
\end{equation}
 In \cite{maxim}, \cite{nora} it has been shown that
loops, which are regions where ($\kappa_i, \tau_i$)  are variable, can be constructed  in terms
of the dark soliton solution of the generalized discrete nonlinear Schr\"odinger equation that derives
from the energy (\ref{Ekappa}),
\begin{equation}
\kappa_{i+1} = 2\kappa_i - \kappa_{i-1} + \frac{ d V[\kappa]}{d\kappa_i^2} \kappa_i  \ \ \ \ \ (i=1,...,N)
\label{nlse}
\end{equation}
where we set $\kappa_0 = \kappa_{N+1}=0$.  This is the {\it Master equation} that we use to compute the
shape of a folded protein C$_\alpha$ backbone.

%
%
%
%
%
%
%
%
%
%
%
%
%
%
%
%
%
%
%
%
%
%
%
%
%
%
%
%
%
%

\subsection{D: Soliton Ansatz}

We do not know the explicit form of the dark soliton solution to the present discrete nonlinear 
Schr\"odinger equation. But a numerical approximation can be easily constructed
using the procedure described in \cite{nora}. 
Furthermore, since it turns out that in the case of proteins 
the first term in (\ref{Vkappa}) is small, 
an excellent approximation \cite{peng}
is given  by the {\it naive} discretization of the 
continuum dark NLSE soliton  \cite{faddeev},
\begin{equation}
\kappa_i  =  \frac{ 
(\mu_{1} + 2\pi N_1)  \cdot e^{ \sigma_{1} ( i-s)  } - (\mu_{2} + 2\pi N_2) \cdot e^{ - \sigma_{2} ( i-s)}  }
{e^{ \sigma_{1} ( i-s) } +  e^{ - \sigma_{2} ( i-s)}   }
\label{bond2}
\end{equation}
Here $s$ is a parameter that determines the backbone site location of the soliton center. This is
the center of the fundamental loop that we describe by 
the soliton.  The $\mu_{1,2} \in [0,\pi] $ are parameters. 
In the case of proteins the values of $\mu_{1,2}$ are entirely determined by the adjacent helices and strands.
Far away from the soliton center we  have
\[
\kappa_i \ \to \left\{ \begin{matrix}  \mu_1 \  & \mod (2\pi) \ \ \ \ i > s \\  -\mu_2 \  & \mod(2\pi) \ \ \ \ i < s
\end{matrix} \right.
\]
For $\alpha$-helices and $\beta$-strands the $\mu_{1,2}$ values are determined by (\ref{bc1}), (\ref{bc2}).
Negative values of $\kappa_i$ are related to the positive values by (\ref{dsgau}).

The $N_1$ and $N_2$ constitute the integer parts 
of $\mu_{1,2}$ and for simplicity we shall take $N_1 = N_2 \equiv N$. This integer is 
like a covering number, it determines  how many times $\kappa_i$ covers  the fundamental 
domain $[0, \pi]$  when we traverse the soliton once. We introduce this integer for the following reason: 
The Ansatz (\ref{bond2}) is monotonic but in general the values of $\kappa_i \in [0,\pi]$ that we obtain
from PDB are not. Whenever we encounter a site $i$ where $\kappa_i$ in the PDB data fails to be
monotonic, we either add or subtract $2\pi$ to its value, to regain a monotonic data profile.
This shift does not have any 
effect on the backbone geometry.
In this manner we utilize the multi-valuedness of $\kappa_i$ to convert any sequence $\{ \kappa_i \}$
into a monotonic one, that we can then approximate by the Ansatz (\ref{bond2}).

Note that for $\mu_1 = \mu_2$  and $\sigma_1 = \sigma_2$ we  recover the hyperbolic tangent. In this
case the two regular secondary structures before and after the loop are the same.
Moreover,  {\it only}  the (positive)  $\sigma_1$ and  $\sigma_2$ are 
intrinsically loop specific parameters. 
They specify the length of the loop and as in the case of the $\mu_{1,2}$
they are combinations of the parameters in (\ref{E1}).

Similarly, in the case of the torsion angle there is only 
one loop specific parameter in (\ref{tauk}). The overall, common
scale of the four parameters is again irrelevant 
and two of the remaining three parameters are fixed by  the regular
secondary structures that are adjacent to the loop. 

Long protein loops, and entire super-secondary protein structures such as a helix-loop-helix 
can be constructed by combining together solitons  (\ref{nlse}), (\ref{tauk}). A typical 
short super-secondary structure that is described by a single soliton 
involves at least 15-20 different amino acids, often even more. As a consequence 
our DNLS equations and  the explicit soliton
Ansatz compute the 45-60 spatial coordinates of the ensuing C$_\alpha$ 
carbons in terms of of the five essential universal parameters in (\ref{E1}).  This implies 
that the DNLS equations comprises a highly under-determined system of equations, and the
key physical principles of our approach are experimentally testable.

In \cite{peng2} it has been shown 
using the Ansatz (\ref{bond2}) that over 92$\%$ of PDB configurations can be constructed in terms of 200 explicit soliton profiles. This makes a strong case that  the solitons of the DNLS equation are  
the modular building blocks of folded proteins \cite{peng2}. 

%
%
%
%
%
%
%
%
%
%
%
%
%
%
%

\subsection{E: Side-chain energy function}

In order to account for the C$_\beta$ contribution to the protein free energy, we
augment (\ref{E1})  by terms that involve the variables  
($\theta_i, \varphi_i$) in (\ref{uvec}). We shall assume  that side-chain directions 
are {\it locally slaved} to the backbone.  By an explicit analysis of 
PDB structures using Frenet frames  this can  be confirmed to be the case 
\cite{fan}, as shown in Figure 2. 

The C$_\beta$ latitude angles $\theta_i$ are gauge {\it i.e.}  frame invariant: The $\theta_i$
are entirely determined by the tangent vectors $\mathbf t_i$, and consequently can not depend on the
choice of framing.
To the leading order we may then assume that  each $\theta_i$ interacts  locally, with
the corresponding $\kappa_i$ only. 
The leading order contribution is obtained by Taylor expanding a general interaction potential
around the ($\kappa_i$ dependent) equilibrium values of the $\theta_i$,
\begin{equation}
E_\theta  =
\sum\limits_{i=1}^N
\left\{  \frac{d_\theta}{2} \, \kappa_i^2 \theta_i^2  -  b_\theta \kappa_i^2  
\theta_i - a_\theta \theta_i   +  \frac{c_\theta}{2}  \theta^2_i 
\right\}   + \dots
\label{E2a}
\end{equation}
where the additional terms are of higher order in powers of $\kappa_i$ and $\theta_i$.
From this we solve for $\theta_i$,
\begin{equation}
\theta_i = \frac{a_\theta + b_\theta \kappa_i^2}{c_\theta + d_\theta \kappa_i^2} 
\label{thei}
\end{equation}
Again, as in the case of (\ref{tauk}),   the overall scale cancels which leaves us
with only three independent parameters.  As visible from Figure 2, 
the fluctuations in $\theta_i$ around the
average value (\ref{thetave}) are small. From this Figure we also learn \cite{fan} that 
these fluctuations are slaved to the backbone geometry, 
which is dictated by the $\kappa_i$. This confirms that  the present 
approximation (\ref{E2a}) is reasonable.

Unlike $\theta_i$, the C$_\beta$ longitude angle $\varphi_i$ does not remain intact under the
frame rotation (\ref{discso2}) but transforms according to (\ref{varp1}). Consequently it can form
a SO(2) frame ({\it i.e.} gauge) invariant combination with the backbone torsion angle (\ref{tors}). Two 
examples of gauge invariant combinations are
\[
\tau_i - \varphi_{i-1} + \varphi_i
\]
and 
\begin{equation}
\varphi_i + \sum_{k=1}^i \tau_k
\label{ginv}
\end{equation}
We prefer to proceed with the second one, it is local in $\varphi_i$. (The first is a difference of the second.)
As in (\ref{E1}) we specify the unitary gauge, which amounts to selecting the Frenet framing along
the backbone.

As visible in Figure 2, the fluctuations in $\varphi_i$ are about as small as those in $\theta_i$.
Moreover, in the combinations (\ref{ginv}) the torsion angles $\tau_i$ are determined locally 
by the bond 
angles according to (\ref{tauk}).  Consequently we may also Taylor expand the $\varphi_i$ 
contribution to the energy,  and following
(\ref{E2a}) we conclude that the leading order contribution is of the form
\begin{equation}
E_\varphi  =  \sum\limits_{i=1}^N
\left\{  \frac{d_\varphi}{2} \, \kappa_i^2 \varphi_i^2  -  b_\varphi \kappa_i^2  \varphi_i - 
a_\varphi \varphi_i   +  \frac{c_\varphi}{2}  \varphi^2_i 
\right\} +\dots
\label{E2b}
\end{equation}
This slaves  $\varphi_i$ to the backbone $\kappa_i$ according to
\begin{equation}
\varphi_i = \frac{a_\varphi + b_\varphi \kappa_i^2}{c_\varphi + d_\varphi \kappa_i^2}
\label{phii}
\end{equation}
Again only three of the four parameters are independent, the 
overall scale drops out.

%
%
%
%
%
%
%
%
%
%
%
%
%
%
%

\subsection{F: Total energy}

We combine (\ref{E1}), (\ref{E2a}) and (\ref{E2b}) to arrive at the total energy
\begin{equation}
E = E_\kappa + E_\tau + E_\theta + E_\varphi 
\label{E0}
\end{equation}
\begin{equation}
= - \sum\limits_{i=1}^{N-1}  2\, \kappa_{i+1} \kappa_{i}  + \sum\limits_{i=1}^N
\biggl\{  2 \kappa_{i}^2 + q\cdot (\kappa_{i}^2 - m^2)^2  \biggr\}
\label{EA}
\end{equation}
\begin{equation}
+ \sum\limits_{i=1}^N \biggl\{ \frac{d_\tau}{2} \, \kappa_{i}^2 \tau_{i}^2   - {b_\tau}\kappa_i^2 \tau_i
- a_\tau  \tau_{i}   +  \frac{c_\tau}{2}  \tau^2_{i} \biggr\}
\label{EB}
\end{equation}
\begin{equation}
+ \sum\limits_{i=1}^N
\left\{  \frac{d_\theta}{2} \, \kappa_i^2 \theta_i^2  -  b_\theta \kappa_i^2  
\theta_i - a_\theta \theta_i   +  \frac{c_\theta}{2}  \theta^2_i 
\right\}   
\label{EC}
\end{equation}
\begin{equation}
+ \sum\limits_{i=1}^N
\left\{  \frac{d_\varphi}{2} \, \kappa_i^2 \varphi_i^2  -  b_\varphi \kappa_i^2  \varphi_i - 
a_\varphi \varphi_i   +  \frac{c_\varphi}{2}  \varphi^2_i 
\right\}  \ + \dots
\label{ED}
\end{equation}
Since the variations  in ($\theta_i, \varphi_i$) are much smaller than those in $\tau_i$, the
ensuing contributions $E_\theta$ and $E_\varphi$ are also much smaller than $E_\tau$. 
Furthermore, according to (\ref{ED})  the backbone torsion angles $\tau_i$ and 
the side-chain angles ($\theta_i, \varphi_i$) are all
slaved to the backbone bond angles $\kappa_i$. As a consequence 
the DNLS equation (\ref{nlse})  is the {\it Master Equation} that entirely determines the 
geometry of a folded protein:
The
C$_\alpha$$-$C$_\beta$ backbone is constructed by first solving for $\kappa_i$. 
The remaining two angles ($\theta_i, \varphi_i$) are then constructed in terms of the $\kappa_i$ 
using (\ref{tauk}), (\ref{thei}) and
(\ref{phii}). 

%
%
%
%
%
%
%
%
%
%
%
%
%
%
%

\subsection{G: Parameters}

The energy function (\ref{E0})-(\ref{ED}) involves a number of parameters. Eventually, we would like
to compute their numerical values directly from the amino acid sequence. But at the moment this has
not yet been achieved.

{\it A priori} it appears that the number of parameters needed  in (\ref{E0})-(\ref{ED}) to describe an entire protein backbone, could be quite large. However, due to the presence of the dark soliton
the number of parameters is actually remarkably small: For each
super-secondary structure such as a helix-loop-helix, whenever the loop can be described in
terms of a single soliton solution to (\ref{nlse}), the potential (\ref{Vkappa}) has  only
four independent parameter combinations. In addition of $q$ and $m$ that characterize the
second and third term, there are only two essential parameters in the first term. Three of these four
parameters can be given the following  interpretations. Two of them determine 
the values of $\kappa_i$ in the ground states such as  (\ref{bc1}), (\ref{bc2}) 
that are located along the backbone 
before and after the soliton {\it i.e.} the type of the helix 
that precedes and follows the soliton.  
The third parameter determines the length of the loop. 
The fourth parameter can then be included as one of the three independent parameters in the 
torsion profile (\ref{tauk}). It can be attributed to the length of the soliton, in terms of torsion.
In addition, in the soliton profile of $\kappa_i$
there is the parameter that specifies the position of the soliton along the backbone. This is
an additional parameter that emerges from the periodicity of the lattice structure (lattice
translation invariance).

Since the overall scale in  (\ref{tauk}) cancels out, the two remaining 
parameter combinations in addition of the loop length, 
become determined by the values of $\tau_i$ 
in the ground states surrounding the soliton {\it i.e.} the type of 
the helix as in (\ref{bc1}), (\ref{bc2}). 

In this way we arrive at the conclusion 
that for the backbone, the only loop specific parameters are those that determine
the lengths of the solitons. All additional  parameters in the energy function
determine the regular secondary structures  such as (\ref{bc1}) and (\ref{bc2}). 
The profiles of all loops are completely fixed by the {\it unique} dark solution solution to 
(\ref{nlse}).

Similarly, we conclude from (\ref{thei}), (\ref{phii}) that in the equations that determine
the Frenet frame orientations of the
C$_\beta$ carbons, there is only one loop specific parameter in both $\theta_i$ and 
$\varphi_i$. In each equation, the overall scale factor cancels out
and the values of the
additional  two independent parameters are fully specified by the 
regular secondary structures adjacent to the loop.

Since (\ref{E0})-(\ref{ED}) aims to predict the 45-60 space coordinate of C$_\alpha$, and the corresponding
30-45 directional coordinates of the C$_\beta$ in terms of 11 essential parameters, we have 
a highly underdetermined system of equations. 
This implies  that the physical principles of our approach 
are experimentally testable.

In \cite{peng2} we have found that most crystallographic protein structures 
in PDB can be described in a modular fashion and with experimental B-factor precision, 
by combining together no more than 200 explicit soliton profiles.  We propose that by
learning how to compute the parameter values directly from the sequence, 
the geometric shape of most
folded proteins can be constructed simply by solving the Master equation (\ref{nlse}). 

%
%
%
%
%
%
%
%
%
%
%
%
%
%
%

\subsection{H: Fluctuations around solitons}

As such, the equations (\ref{nlse}), (\ref{tauk}), (\ref{thei}), (\ref{phii}) describe the  
critical points of the energy function (\ref{E0}). This energy function should be duly interpreted as 
describing the effective
long distance limit of the full microscopic second-quantized Schr\"odinger operator. 
As such, (\ref{E0}) then relates to proteins in the sense of a semi-classical approach. 
This kind of semi-classical description is common in quantum field theories. 
There, it is often boldly used to describe phenomena at length scales that are several 
orders of magnitude smaller than anything which may have any direct relevance to proteins. 
We now wish to estimate the short distance scale, at which  we expect 
the semiclassical approach to proteins  based on (\ref{E0}), to break down due to 
quantum mechanical zero-point fluctuations.

The backbone profile (\ref{nlse}), (\ref{tauk})
describes the C$_\alpha$  lattice in the limit  where thermal fluctuations vanish. 
But even near zero temperature the protein remains subject to residual zero-point 
fluctuations that can not always  be ignored.  It is difficult to estimate and even 
harder to accurately calculate the 
amplitude of these zero-point fluctuations.  For a realistic order of magnitude
estimate we inspect the distribution of the B-factors that characterize 
experimental uncertainties in PDB data. We use all  those PDB structures where
the crystallographic measurements have been made at temperatures less  than 50K. 
\begin{figure}
  \begin{center}
    \resizebox{7.5cm}{!}{\includegraphics[]{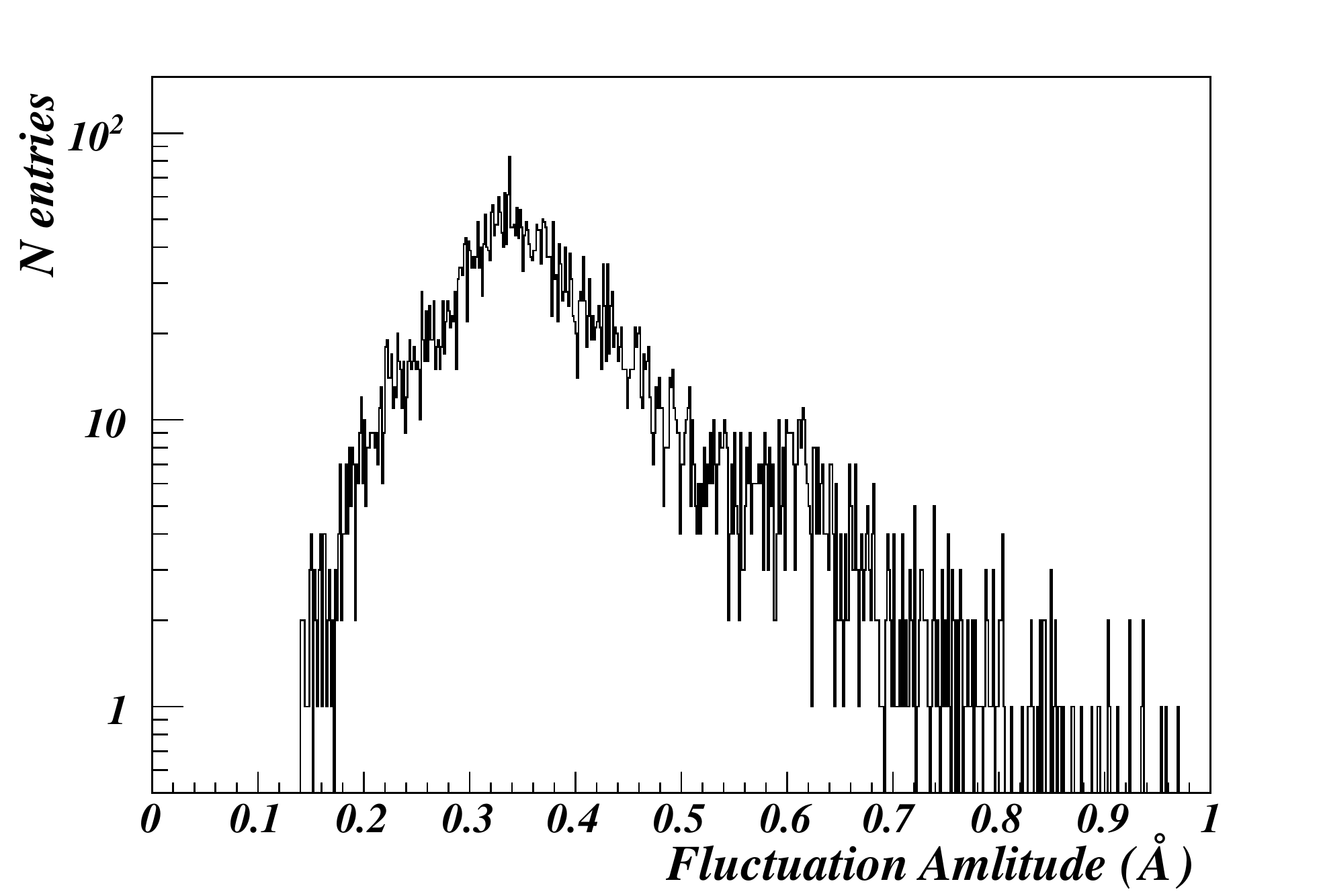}}
    \caption{The number of entries in PDB with temperature below 50K {\it vs.}  Debye-Waller fluctuation distance. Note the logarithmic scale.}
    \label{fig:simple}
  \end{center}
\end{figure}
The result  is displayed in Figure 3. It  shows  that for the C$_\alpha$ carbons 
the zero point fluctuations have a lower bound  which is in the vicinity of  0.15 \AA.   
Consequently we estimate that the precision of our semi-classical approach
can at best be of the same order of magnitude. We account for these zero point  fluctuations by 
dressing our (semi)classical backbone profiles with a tubular dominion that has a radius of 0.15 \AA. 
In particular, this tubular dominion accounts for the bond length fluctuations, around the 
average distance (\ref{dist}) between the neighboring C$_\alpha$ carbons. 
 As can be seen from Figure 1, the fluctuations in the C$_\alpha$$-$ C$_\alpha$ distances 
are normally within range of 0.15 \AA.

%
%
%
%
%
%
%
%
%
%
%
%
%
%
%
%
%
%
%
%
%
%
%
%
%
%
%
%
%
%

\section{III: $\lambda$-repressor protein as a multisoliton configuration}

\subsection{A: Loop spectrum of 1LMB}

We start our soliton-based investigation 
of the $\lambda$-repressor protein by analyzing its $(\kappa_i, \tau_i$) spectrum.
This will identify the putative soliton content.
We use the first chain of the homo-dimer with the PDB code 1LMB.  The 
structure is conventionally interpreted as  a four loop configuration, and
the second loop is the DNA binding one.

From the PDB file we read the C$_\alpha$ coordinates. We  compute the tangent 
vectors from (\ref{t}) and the binormal
vectors from (\ref{b}), and the bond and torsion angles from (\ref{bond}) and (\ref{tors}). 
We construct these angles using the standard convention that $\kappa_i \in [0,\pi]$. 
We locate the regions where
the torsion angles $\tau_i$ are subject to large fluctuations. In these regions we judiciously 
implement the transformation (\ref{dsgau}).  This identifies the soliton structures in the loops.
Both the motivation and the technical details 
of the soliton identification procedure are 
described in \cite{maxim}  and \cite{hu}. 

In the left hand side of Figure 4 we show the ($\kappa_i, \tau_i)$ 
spectra that we obtain from the PDB data using 
the standard differential geometric convention that curvature is positive $\kappa_i >0$.
Each of the four figures describes the spectra over 
one of the four loops of 1LMB, as they are identified in PDB.
\begin{figure}[!hbtp]
  \begin{center}
    \resizebox{8.5cm}{!}{\includegraphics[]{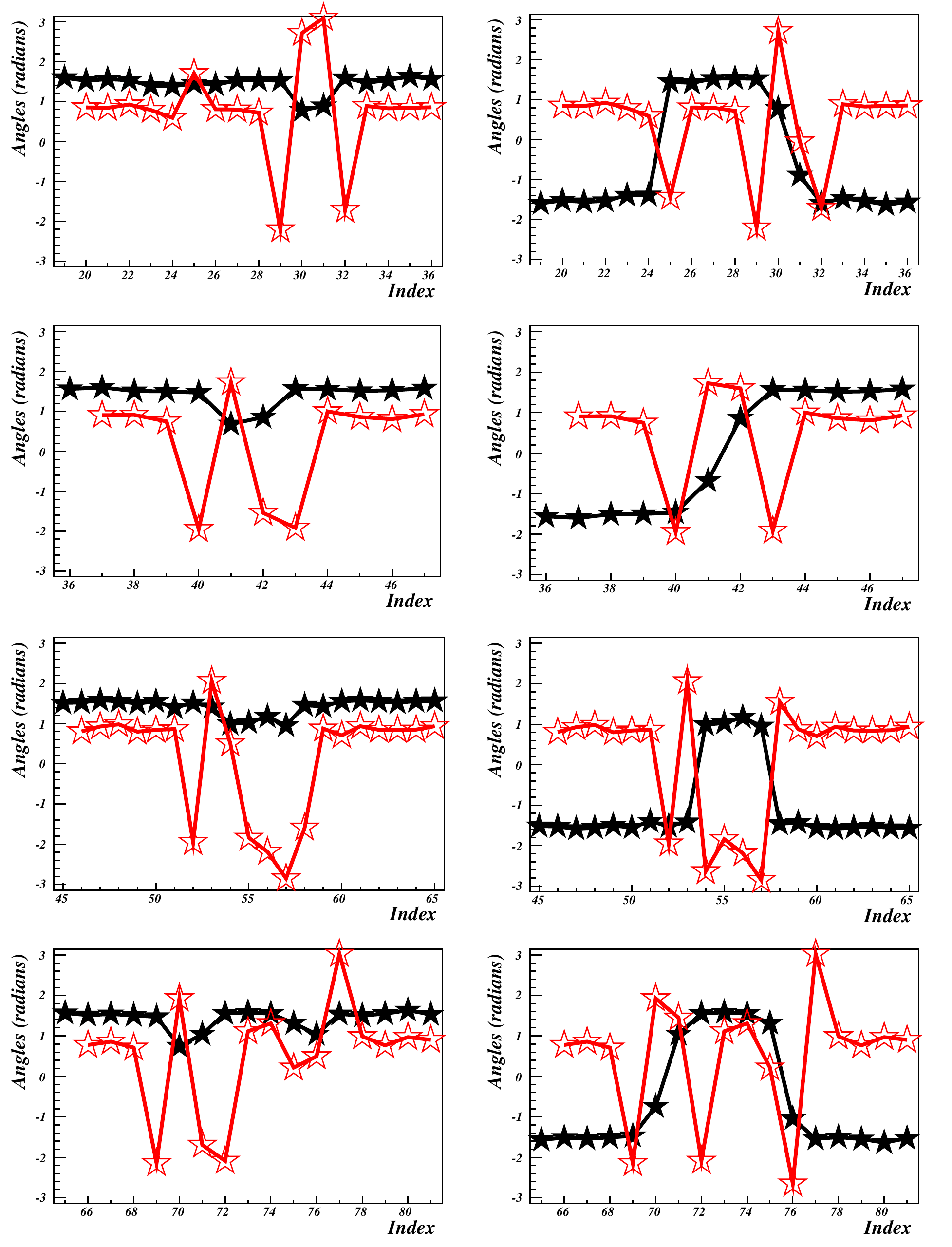}}
    \caption{(Color online)  On the left column, we show the ($\kappa_i, \tau_i$) spectrum for the four PDB
    loop structures of 1LMB, that we compute with the prevailing differential geometric convention
    that curvature is a nonnegative quantity.  Black is the bond angle $\kappa_i$, grey (red) is 
    the torsion angle $\tau_i$.
    On the right we display the corresponding spectra after 
    we have identified the soliton profiles in $\kappa_i$, using (\ref{dsgau}).  All PDB loop structures except 
    for the second, display the
    profile of a soliton pair. Only the second PDB loop can be 
    identified as a single soliton state.}
    \label{fig:simple}
  \end{center}
\end{figure}
We observe that in each Figure 4 on the left hand side, there is a region where
the torsion angle $\tau_i$  fluctuates rapidly 
between positive values and negative values. On general grounds \cite{maxim}, 
\cite{hu} a region where the
torsion angle is solely positive and only subject to small variations is a putative regular 
secondary structure.
On the other hand, a region where the torsion angle  fluctuates between positive and negative values is an 
indicative of an inflection point \cite{hu}, this kind of fluctuation suggests the presence of solitons. 
By judiciously applying the gauge transformation (\ref{dsgau}) in the regions where $\tau_i$ fluctuates
between positive and negative values, we find 
that the $\kappa_i$ profiles in each of the left hand side Figure 4 indeed describe solitons: 
Comparing with (\ref{bc1}), (\ref{bc2}) we conclude that
the first loop structure in the PDB data is a pair of solitons separated 
by a short $\alpha$-helix. The second PDB loop is a single soliton that interpolates between two
$\alpha$ helices. The third loop structure in the PDB data 
is a pair of solitons  separated by a short $\beta$-strand. Finally, the fourth PDB loop is a pair
of solitons, separated by a short $\alpha$-helix. 

Notice that our spectral analysis based structure identification 
is a refinement of the  conventional one, which utilizes techniques such as the
presence or absence of hydrogen bonds  to conclude whether a site
is part of a  regular secondary structure or part of  a loop. 
Consequently very short helical structures that become clearly
visible in our ($\kappa_i, \tau_i$) spectral analysis, can be interpreted differently in 
more conventional approaches.

%
%
%
%
%
%
%
%
%
%
%
%
%
%
%

\subsection{B: Soliton Ansatz }

We proceed to evaluate the parameters in (\ref{bond2}), (\ref{tauk}) that give our best fit to those
seven soliton profiles, identified by analyzing the ($\kappa_i, \tau_i$) spectrum. 
In Table 1
\begin{table}[tbh]
\begin{center}
\caption{Our best parameter values for each of 
the seven solitons in Figure 2.  Note that $m_1$, $m_2$ are determined $\mod(2\pi)$.}
\vspace{3mm}
\begin{tabular}{c|cccc}

Soliton & $c_1$ & $c_2$ & $m_1$ & $m_2$  \\
\hline  \\ [-3mm]
(8,30) & ~ 2.00441 ~ & ~ 1.99595 ~ & ~ 26.65124 ~ & ~ 26.68412 ~   \\
\hline \\ [-3mm]
(26,39) & 2.94889 & 2.95201 & 70.67882 & 70.60369  \\
\hline \\ [-3mm]
(36,47) & 2.89729 & 2.90755 & 39.27387 & 39.22546  \\
\hline \\ [-3mm]
(46,59) & 2.97927 & 3.00015 & 1.07948 & 1.52942  \\
\hline  \\ [-3mm]
(56,65) & 2.96486 & 2.97087 & 26.69087 & 26.25280  \\
\hline \\ [-3mm]
(62,74) & 2.94948 & 2.94547 & 20.43071 & 20.38220  \\
\hline \\ [-3mm]
(74,87)& 2.89725 & 2.89945 & 89.50870 & 89.55252  \\
\hline \end{tabular}
\vskip 0.4cm
\begin{tabular}{c|ccc}

Soliton & s & a/b & d/b  \\ 
\hline \\ [-3mm]
(8,30)&  ~ 24.50259 & ~ $-9.9921 \cdot 10^{-2}$ ~ & $4.2191 \cdot 10^{-5}$ ~ \\
\hline \\ [-3mm]
(26,39)& \ 30.49642 & $-1.5114 \cdot 10^{-7} $ & $ 1.0662 \cdot 10^{-11} $ \\
\hline \\ [-3mm]
(36,47) &  41.39325 & $-5.3794 \cdot 10^{-7}$ & $7.4566 \cdot 10^{-11}$ \\
\hline \\ [-3mm]
(46,59) & 53.67225 & $  5.1477 \cdot 10^{-7}$ & $\! -5.1529 \cdot 10^{-7} $  \\
\hline \\ [-3mm]
(56,65) &  57.85123 & $-9.62942 \cdot 10^{-8}$ & $1.45097 \cdot 10^{-12}$ \\
\hline \\ [-3mm]
(62,74) &  70.22069 & $-9.27151 \cdot 10^{-7}$ & $3.05202 \cdot 10^{-10}$ \\
\hline \\ [-3mm]
(74,87) &  75.56315 & $-7.13705 \cdot 10^{-7}$ & $1.8457 \cdot 10^{-11}$ \\
\hline 
\end{tabular}
\end{center}
\label{solenoid}
\end{table}
we present our best parameter values. The parameters can be computed
by various different techniques. Here we have chosen a Monte Carlo search 
that is fast and gives very good accuracy. In Table 1 we also 
identify the corresponding super-secondary structures by their PDB backbone
indices. Note that since our approach 
is based on the spectral analysis of bond and torsion angles  and since the definition of a bond angle involves three sites while that  of the  torsion angle involves four,  three residue indices at both ends of the 1LMB chain are absent in the ($\kappa_i,\tau_i$) list. 
Notice also that we have introduced some overlap between the different super-secondary structures. This ensures that we can eventually combine them 
together into the full 1LMB backbone.
Moreover, in the case of  all except the fourth soliton, we have utilized the multi-valuedness 
in the definition of the bond angle to extend the range of its values outside of the fundamental domain $\kappa_i \in [0,\pi]$.  This corresponds to selecting non-vanishing integers $N_1$, $N_2$ in the 
Ansatz (\ref{bond2}).  For simplicity we have limited our Monte Carlo search to the symmetric 
case $N_1=N_2$, but for better accuracy the soliton profiles could also be given
asymmetric integer parts.  The numerical values of the bond angles $\kappa_i$ 
that we compute from (\ref{bond2}) 
using the parameter
values in Table I, are to be reduced onto the fundamental domain 
$[0,\pi]$ using the $mod(2\pi)$ multivaluedness.    

We recall from Section II D, that the introduction of non-vanishing values of $N_1$ and $N_2$ enables us 
to account for the non-monotonic profile that the bond angles of the PDB configuration display when restricted into the fundamental domain: At each point where the profile of the PDB data becomes non-monotonic, we simply add (or
subtract) $2\pi$ until we obtain a monotonic profile.  In this manner, by allowing  the bond angle to take values over a larger domain,  the $\kappa_i$  profile of the PDB data  over each soliton 
can be made monotonic  which improves the accuracy of the fitted soliton profile (\ref{bond2}). See Figure 5
where we display the second PDB loop together with its approximation by the Ansatz (\ref{bond2})
in the extended range, as an example. 
 \begin{figure}[!hbtp]
  \begin{center}
    \resizebox{8.cm}{!}{\includegraphics[]{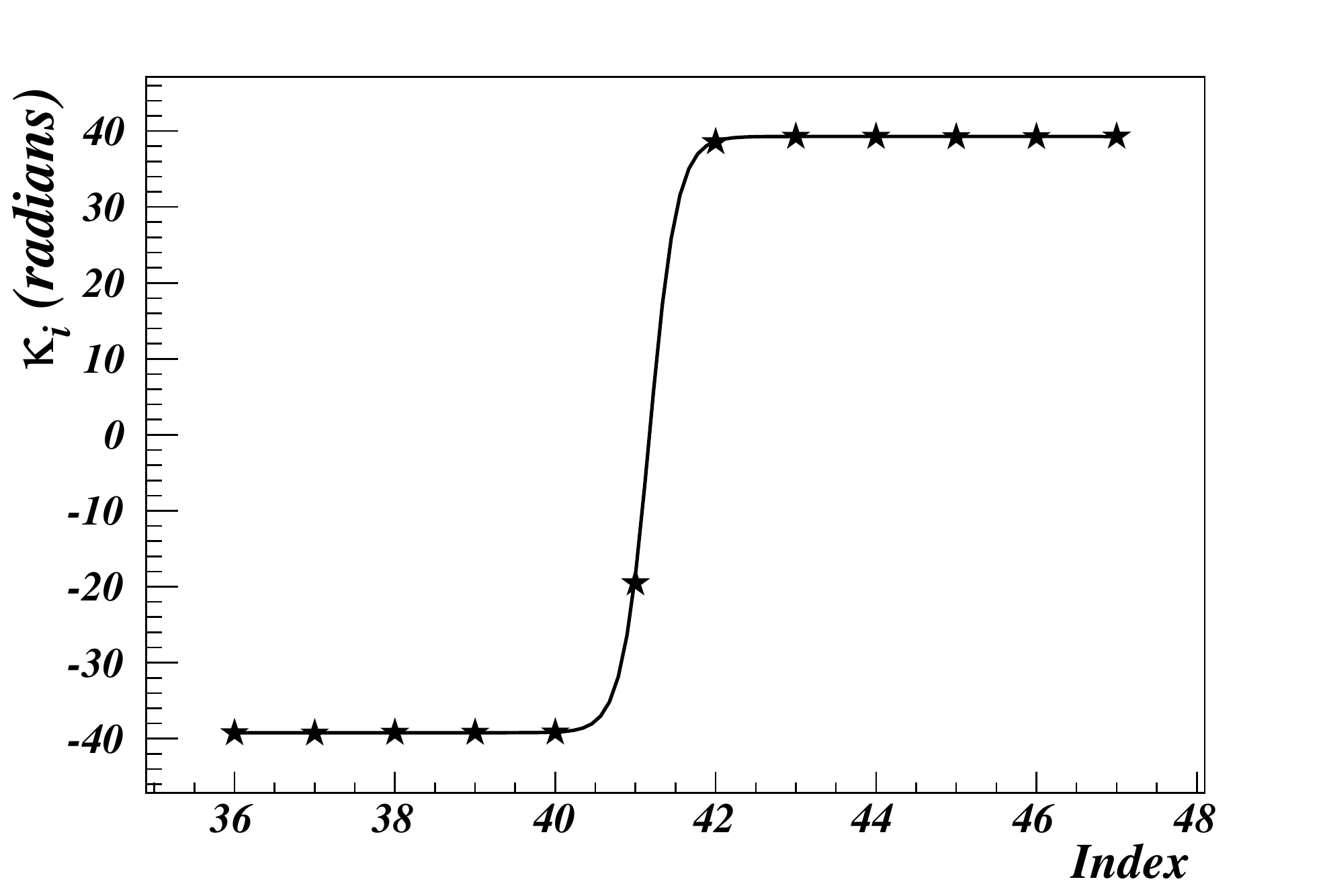}}
    \caption{The data points correspond to the second loop of 1LMB, after we have properly
    extended the range of the bond angles, by using the inherent multivaluedness of these angles. The interpolating
    function describes the Ansatz (\ref{bond2}) with the parameter values given in Table I.
    }
    \label{fig:simple}
  \end{center}
\end{figure}

We compute the torsion angles $\tau_i$  from (\ref{tauk}), before implementing the $2\pi$ reduction of the
bond angles.  We then reduce the ensuing values of the $\tau_i$ into the fundamental domain
$\tau_i \in (-\pi, \pi]$ using the $2\pi$ periodicity. The underlying multivaluedness entirely accounts for the 
fluctuations in the $\tau_i$ profile.
\begin{figure}[!hbtp]
  \begin{center}
    \resizebox{8.cm}{!}{\includegraphics[]{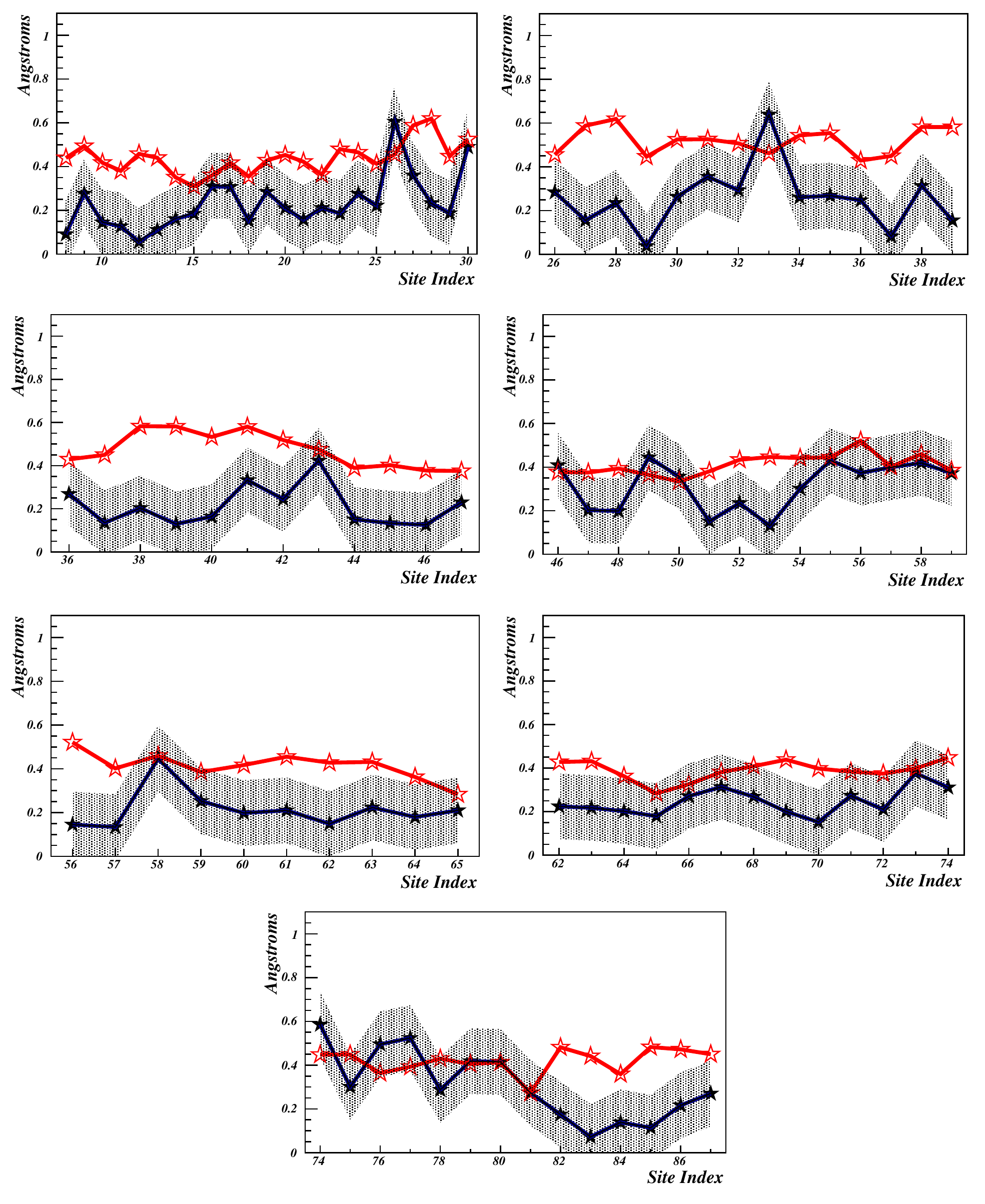}}
    \caption{(Color online)  The distance between the PDB backbone of the first 1LMB chain and its approximation by
    the seven solitons (\ref{bond2}), (\ref{tauk}) as a function of the residue number. 
    The black line denotes the distance between  the soliton
    and the corresponding PDB configuration, the grey area around the black line 
    describes  the estimated 0.15 \AA~  zero point fluctuation distance around solitons, obtained from
    Figure 3. 
    The grey (red) line denotes the Debye-Waller distance that is computed form the B-factors in PDB.  
    }
    \label{fig:simple}
  \end{center}
\end{figure}

In Figure 6 we compare the explicit backbone profiles that we have computed 
from (\ref{bond2}), (\ref{tauk}), (\ref{DFE2}), (\ref{dffe}) using the parameter 
values in Table I, with the 1LMB data in PDB. We find that for each soliton, our backbone
profiles describe the  structural motifs of 1LMB with a precision that  is {\it substantially} better 
than the experimental  precision which is determined by B-factors. This persist even when we account 
for the 0.15 \AA~ estimate of the zero point fluctuations around
our solitons.   With these highly precise soliton profiles we  can {\it unambiguously} 
conclude that 1LMB has  a total of  seven solitons. Two of the  $\alpha$-helices and the 
sole $\beta$-strand are so short that the ensuing soliton pairs  are  interpreted as 
single loops in the conventional, hydrogen bond based analysis of the PDB data.  
The only motif where our construction  identifies a single PDB loop as a single isolated soliton, 
is the DNA binding one.  All of the remaining three loops consist of a pair of solitons  with profile
(\ref{bond2}),  (\ref{tauk})  that are separated from each other either 
by a {\it very} short $\alpha$-helix in case of the residues  ($23,33$) and ($69,90$), or  by 
a {\it very} short $\beta$-strand in case of  
residues ($51,61$). 

We have performed a statistical analysis on the occurrence of our seven solitons in all PDB proteins. 
In Table 2 
\begin{table}[!htb]
\begin{center}
\caption{ The (minimal) soliton sites that we have  used in searching for matching 
structures in PDB, together with the number of matches. 
The search is limited to those x-ray structures that have a resolution better than 2.0 \AA ~and a match is a configuration that 
deviates less than 0.5 \AA ~in total root mean square distance (RMSD) from the soliton.} 
\vspace{3mm}
\begin{tabular}{|c|cccc|}
\hline
Soliton & 1 & 2 & 3 & 4  \\
\hline
Sites   & ~(20,28)  ~& ~(27,36)  ~& ~(36,46)  ~& ~(50,58)  ~ \\
\hline 
Matches  & 9601  & 4  & 810  & 159   \\  
\hline 
\end{tabular}
\vskip 0.4cm
\begin{tabular}{|c|ccc|}
\hline
Soliton & 5 & 6 & 7 \\
\hline
Sites   & ~ (55,63)  ~&~ (66,75) ~ & ~(74,82)~  \\
\hline 
Matches   & 1552  & 1342  & 406  \\  
\hline 
\end{tabular}
\end{center}
\label{table2}
\end{table}
we list the number of matches that each of these solitons has when we search PDB for 
configurations that deviate from the given soliton by an overall  root mean square distance (RMSD) which is less than 0.5 \AA.  
We have chosen this  cut-off value since it is representative of  the Debye-Waller fluctuation distance 
in the experimental 1LMB data; see Figure 6. 
The interesting observation  is that the second soliton of 1LMB 
that  precedes the DNA recognition helix,  is {\it
unique}  to the $\lambda$-repressor protein.   The {\it only} matching structures are located
in the different PDB entries of the $\lambda$-repressor protein itself. 
We also note that for this soliton the B-factors are
slightly higher than for any of the other six solitons along the 1LMB backbone. 

Finally,  by following  \cite{peng} we have combined
the seven solitons together to describe the sites 8-90 (PDB indexing) of the entire 
1LMB backbone. The result
is shown in Figure 7. 
\begin{figure}[!hbtp]
  \begin{center}
    \resizebox{8.cm}{!}{\includegraphics[]{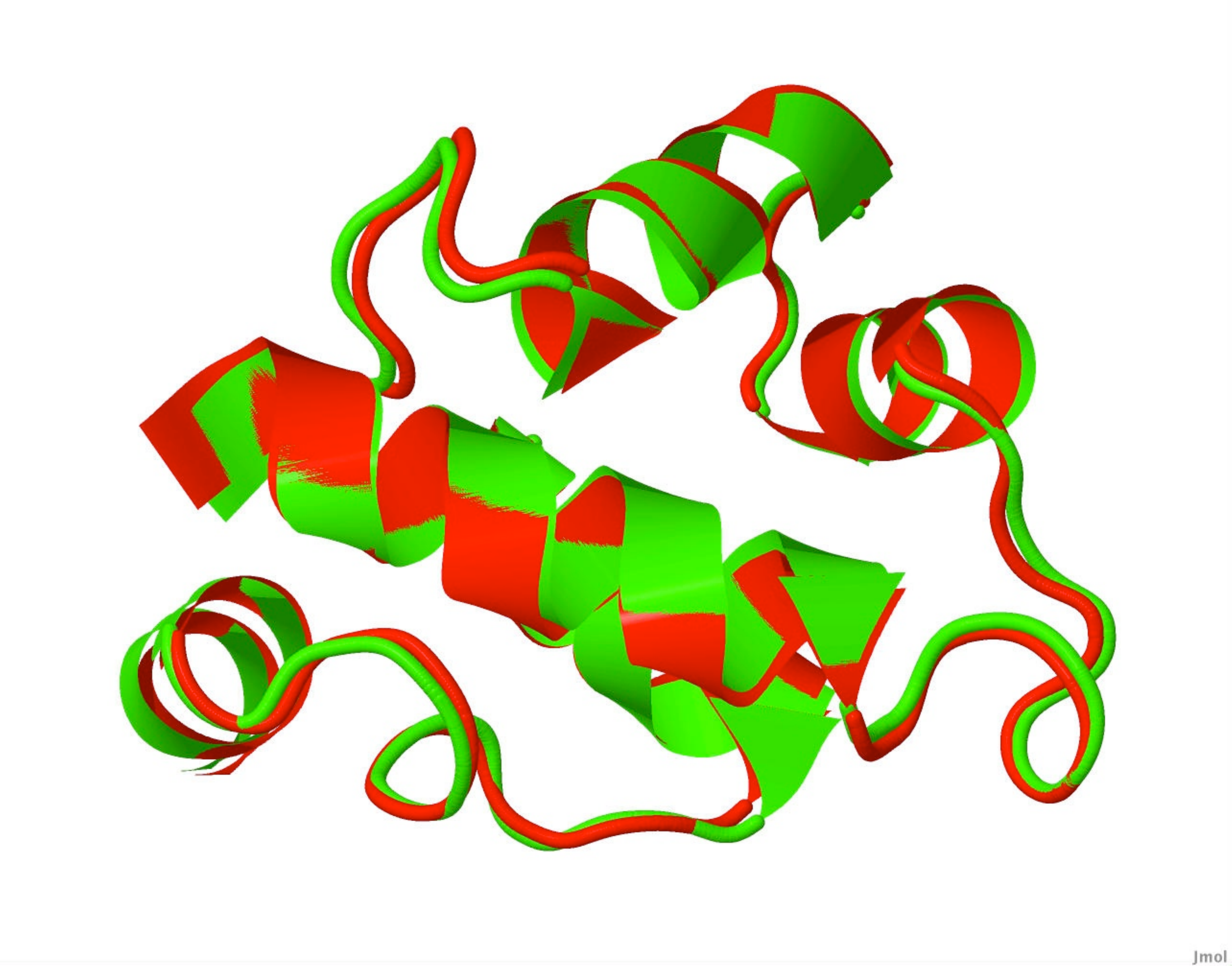}}
    \caption{(Color online)  The PDB backbone configuration of 1LMB (dark red) and its multi-soliton (light green)
    approximation, 
    for sites 8-90. The total RMSD distance is 0.45 \AA.}
    \label{fig:simpl}
  \end{center}
\end{figure}
The overall RMSD accuracy that we get in this manner is 0.45 \AA. This could 
be further improved by adjusting the parameters while combining the solitons. 
But as such, the accuracy displayed by the configuration in Figure 7 is below 
the experimental B-factors. Consequently any improvement is senseless, 
in light of the quality of the experimental data.

%
%
%
%
%
%
%
%
%
%
%
%
%
%
%

\subsection{C: Soliton solution and 1LMB}

We proceed to construct the parameters corresponding to the 
1LMB backbone, for {\it both} the C$_\alpha$-atoms  and the side-chain
C$_\beta$-atoms, using the energy function (\ref{E0}).
We first solve (\ref{nlse}) to obtain the bond angle {\it i.e.} $\kappa_i$-profile. We then construct the 
backbone torsion angles $\tau_i$ and the side-chain ($\theta_i, \varphi_i$) angles from (\ref{tauk}),
(\ref{thei}), (\ref{phii}). We construct a {\it single} solution of (\ref{nlse}), that describes 
the entire chain. 

For simplicity
of construction, we restrict the values of $\kappa_i$ into the  fundamental domain $\kappa_i 
\in [0,\pi]$. As in the case of the Ansatz (\ref{bond2}), a higher precision could  be obtained by 
allowing $\kappa_i$ to  take values beyond the fundamental domain. But with (\ref{E0}), it turns
out that we can reach the B-factor accuracy by constructing the solution of (\ref{nlse}), using the 
fundamental domain only. This is because the solution of (\ref{nlse}) is even better suited 
for modeling proteins than the Ansatz (\ref{bond2}).

To construct the parameters that describe the backbone $\kappa_i$ profile of 1LMB, we
use the iterative learning algorithm of \cite{nora}. It determines 
the parameters by locating the fixed point of 
\begin{equation}
\kappa_i^{(n+1)} \! =  \kappa_i^{(n)} \! - \epsilon \left\{  \kappa_i^{(n)} V'[\kappa_i^{(n)}]  
- (\kappa^{(n)}_{i+1} - 2\kappa^{(n)}_i + \kappa^{(n)}_{i-1})\right\}
\label{ite}
\end{equation}
that minimize the RMSD distance to 1LMB.
Here  $\{\kappa_i^{(n)}\}_{i\in N}$ denotes the $n^{th}$ iteration of an initial 
configuration  $\{\kappa_i^{(0)}\}_{i\in N}$ and $\epsilon$ is some 
sufficiently small but otherwise arbitrary numerical constant. We select 
$\epsilon = 0.01$. A fixed point of (\ref{ite}) clearly satisfies the DNLS equation (\ref{nlse}).
Following \cite{nora} we utilize step-functions to construct an initial configuration for $\kappa_i$.
The ensuing initial profile of $\kappa^{(0)}_i$ is chosen to have the same overall profile
as the properly gauge transformed 1LMB that we display 
in Figure 4 right hand side column. 
A Monte Carlo routine is set up to determine the parameters. For this we 
have developed a  package that we call {\it Propro} \cite{propro}. 
It implements our parameter learning algorithm for a given protein structure,
largely automatizing the entire process. 

In Figure 8 we show a  {\it Propro} screen capture of 
the $\kappa_i$ profile that describes the final multi-soliton solution that yields
the shortest overall RMSD distance between the solution to (\ref{ite}) and 
the 1LMB structure, for the backbone C$_\alpha$ carbons. 
\begin{figure}[!hbtp]
  \begin{center}
    \resizebox{8.5cm}{!}{\includegraphics[]{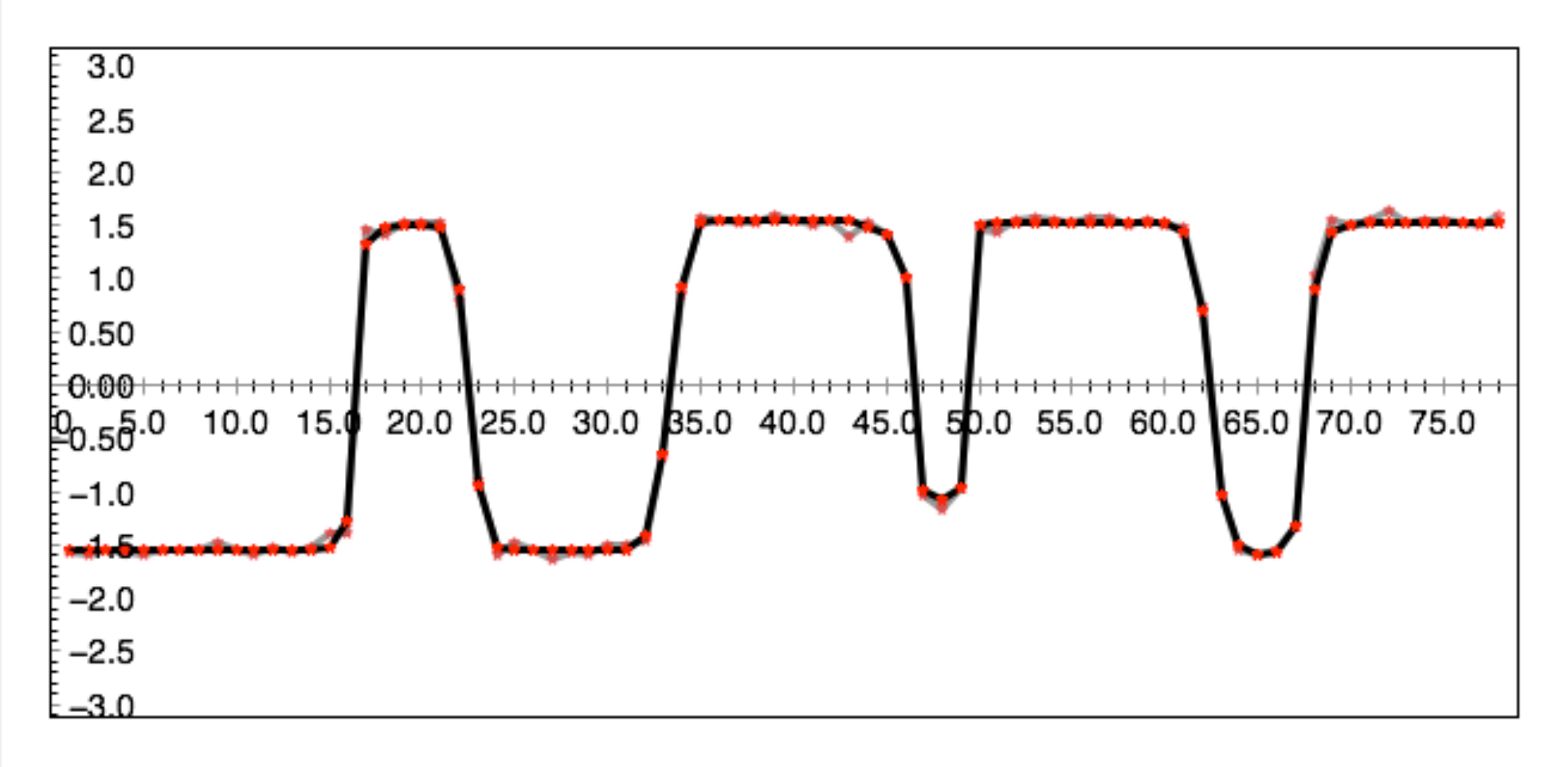}}
    \caption{(Color online)  The multi-soliton solution to (\ref{ite}) (line) and the PDB values of
    1LMB (dots) along the C$_\alpha$ backbone, for the backbone bond angles }
    \label{fig:simple-propro}
  \end{center}
\end{figure}
In Figure 9 we show the corresponding $\tau_i$ profile, computed from (\ref{tauk}).
\begin{figure}[!hbtp]
  \begin{center}
    \resizebox{8.5cm}{!}{\includegraphics[]{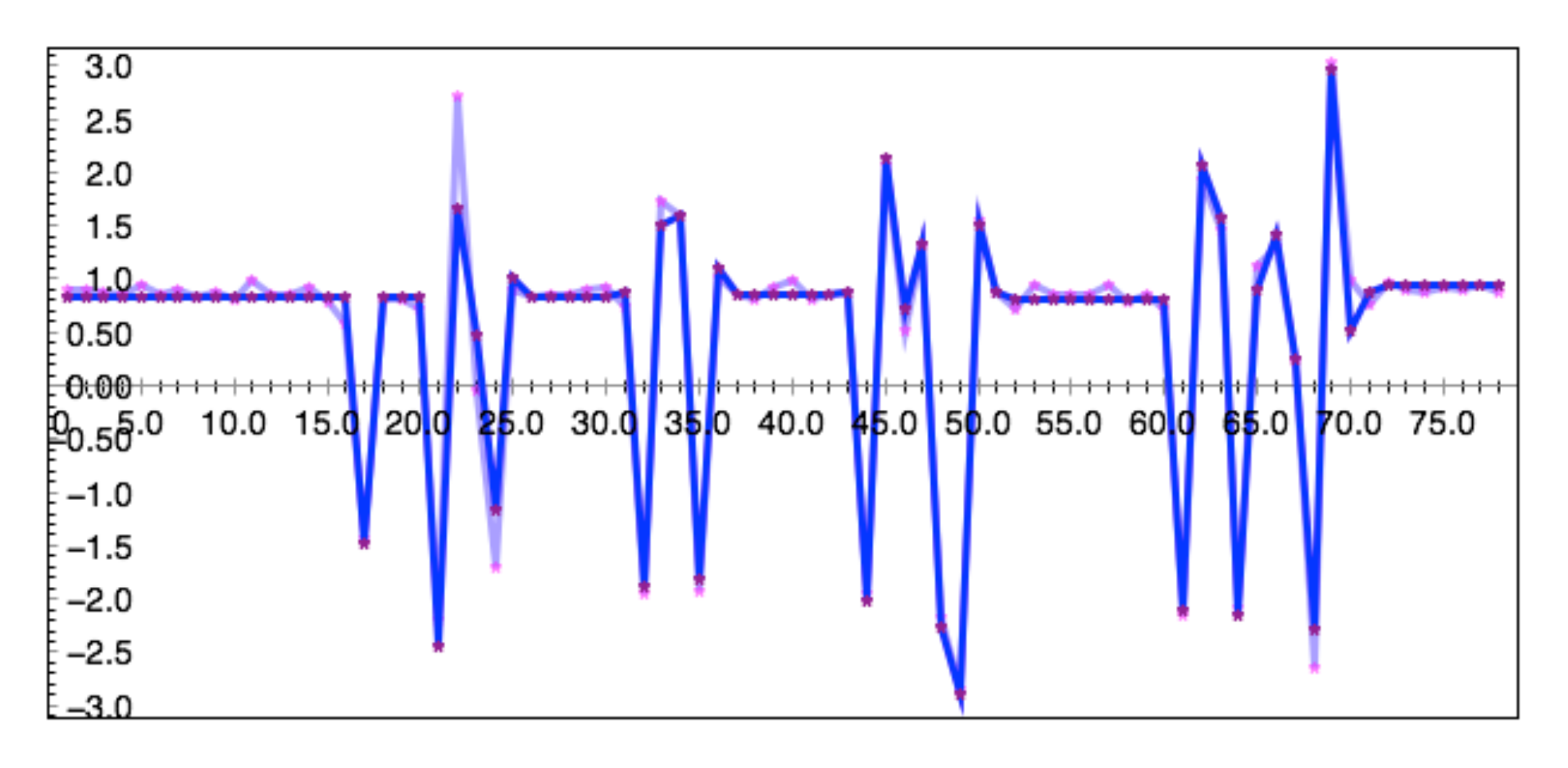}}
    \caption{(Color online)  The profile of torsion angle computed from (\ref{tauk}) (line)  and the corresponding
    PDB values of 1LMB (red dots) along the C$_\alpha$ backbone. }
    \label{fig:simple-propro2}
  \end{center}
\end{figure}
The backbone C$_\alpha$ RMSD distance between out multi-soliton solution and 1LMB
is 0.52 \AA. This is slightly larger that what we obtained with the Ansatz (\ref{bond2}). 
But this time we have not extended the values of $\kappa_i$ outside of the range $\kappa_i \in
[-\pi, \pi]$. In Figure 10 we display the distance between the 1LMB and the soliton 
solution to (\ref{nlse}), (\ref{tauk}) for the individual C$_\alpha$ atoms. For the most part the distance between
our multi-soliton solution and 1LMB is below the Debye-Waller fluctuation distance. The only real
exception is located at the site 31, where the distance between 1LMB C$_\alpha$ carbon
and the soliton solution is close to 1.6 \AA.
\begin{figure}[!hbtp]
  \begin{center}
    \resizebox{8.cm}{!}{\includegraphics[]{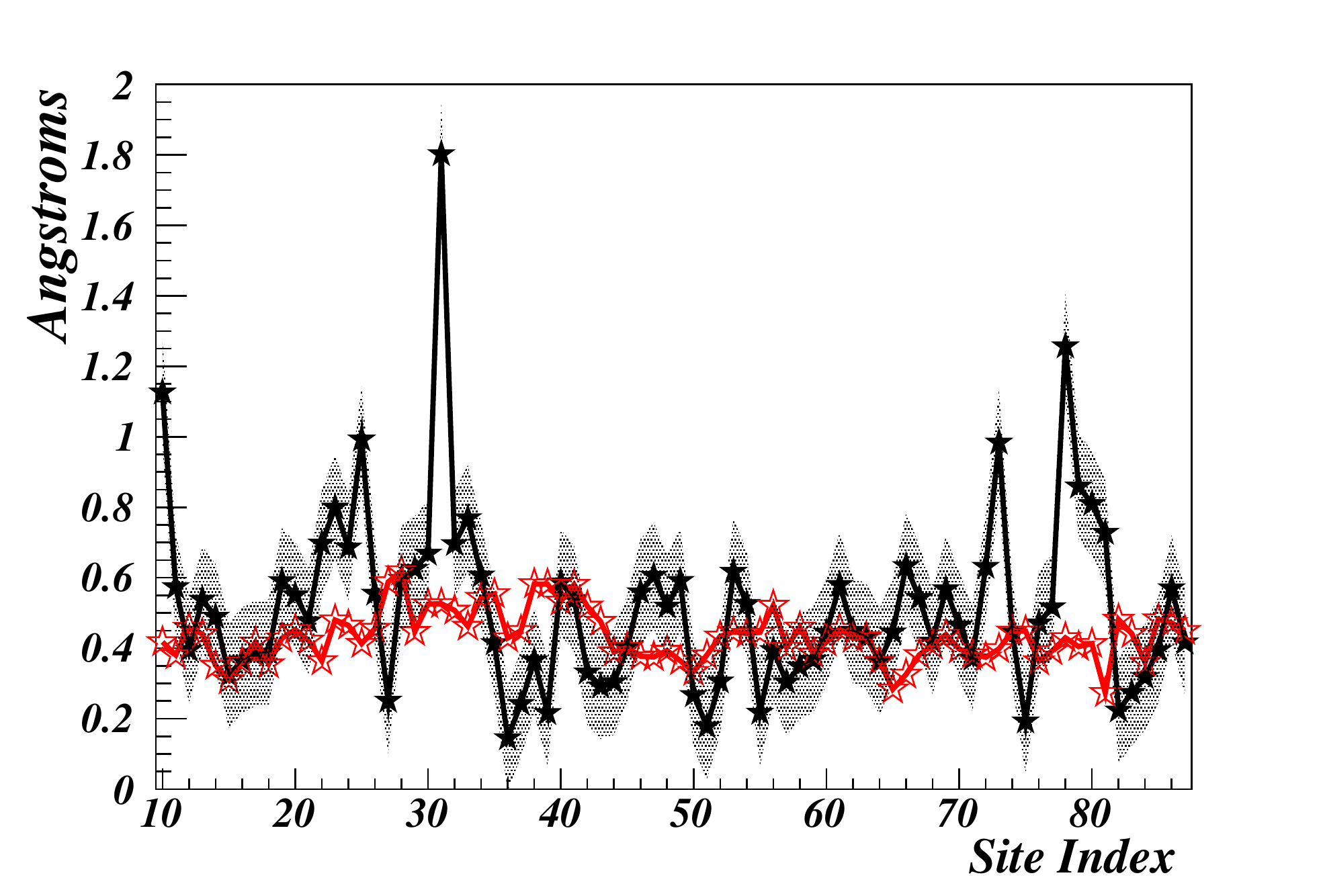}}
    \caption{(Color online)  The distance between the 1LMB backbone and our multi-soliton configuration, constructed
    by solving (\ref{nlse}) (black line). The grey shaded area around the black line describes the estimated 0.15 \AA ~zero point fluctuation distance around the multi-soliton solution. The grey (red) line describes the experimentally measured 
    Debye-Waller fluctuation distance }
    \label{fig:simple}
  \end{center}
\end{figure}
This site is located at the second soliton. We recall that this soliton is unique for 1LMB (see Table II) and
that it was also singled out by the Ansatz (\ref{bond2}). Our analysis indicates that something takes place
at this soliton that warrants a more careful experimental analysis. The properties
of this soliton might have a 
r\^ole in the transition from lysogenic to lytic state.
 
We proceed to extend our multi-soliton to describe both the positions of the C$_\alpha$ carbons,
and the directions of the C$_\beta$ carbons. For this we use (\ref{thei}) and (\ref{phii}). The final
configuration has a combined C$_\alpha$ - C$_\beta$ distance of 0.6 \AA ~ to 1LMB. In Figure 11
we compare the corresponding structures, the 1LMB backbone is translucent.
 \begin{figure}[!hbtp]
  \begin{center}
    \resizebox{8.cm}{!}{\includegraphics[]{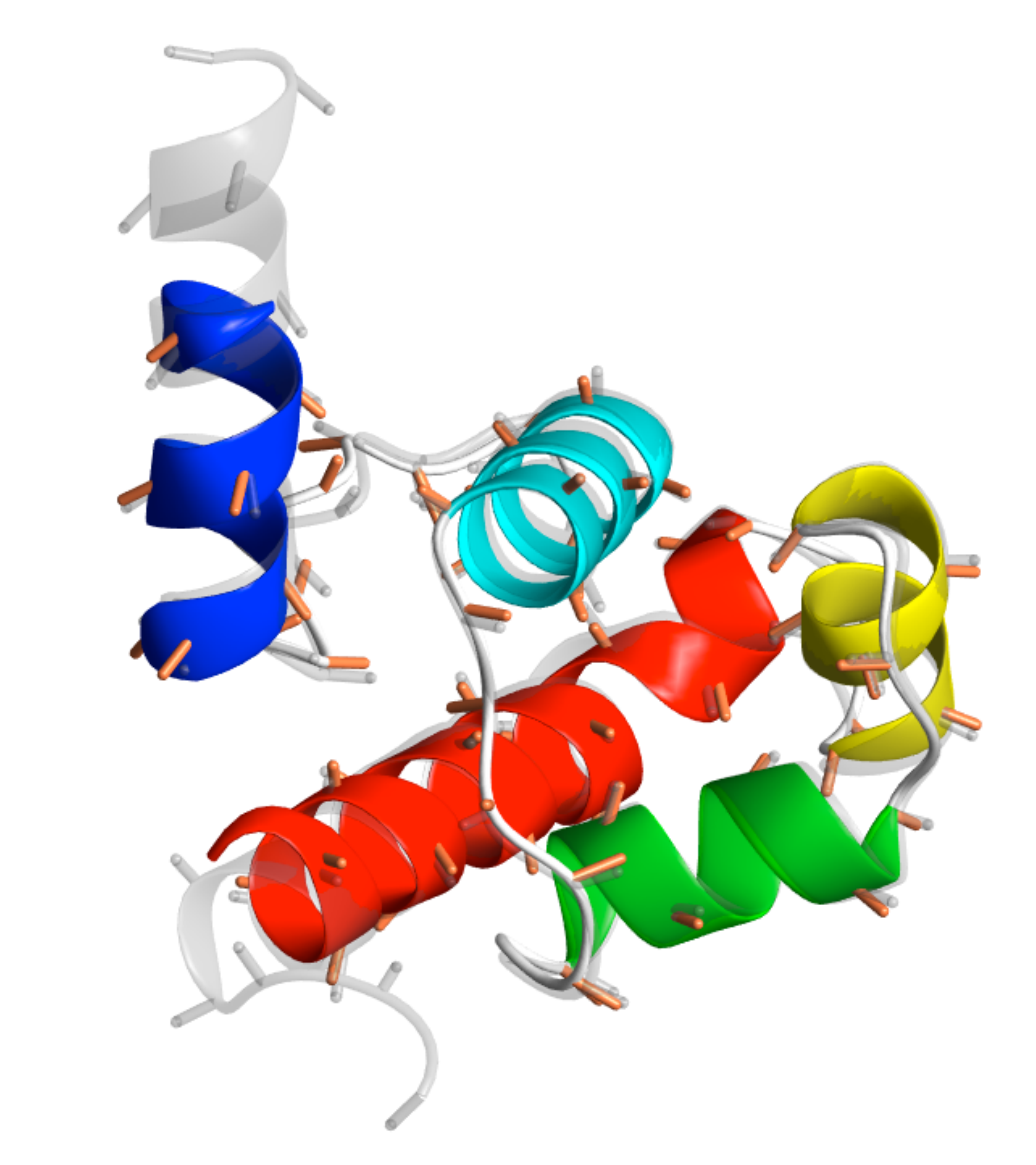}}
    \caption{(Color online)  Comparison between the C$_\alpha$ - C$_\beta$ multi-soliton solution (opaque, with colors 
    online) and the 
    1LMB structure (translucent, grey online). The RMSD distance is 0.6 \AA. }
    \label{fig:simple}
  \end{center}
\end{figure}

In Figure 12 we have a close-up of the region around PDB site 31, where the difference between the
multi-soliton solution and the 1LMB configuration is largest.
\begin{figure}[!hbtp]
  \begin{center}
    \resizebox{8.cm}{!}{\includegraphics[]{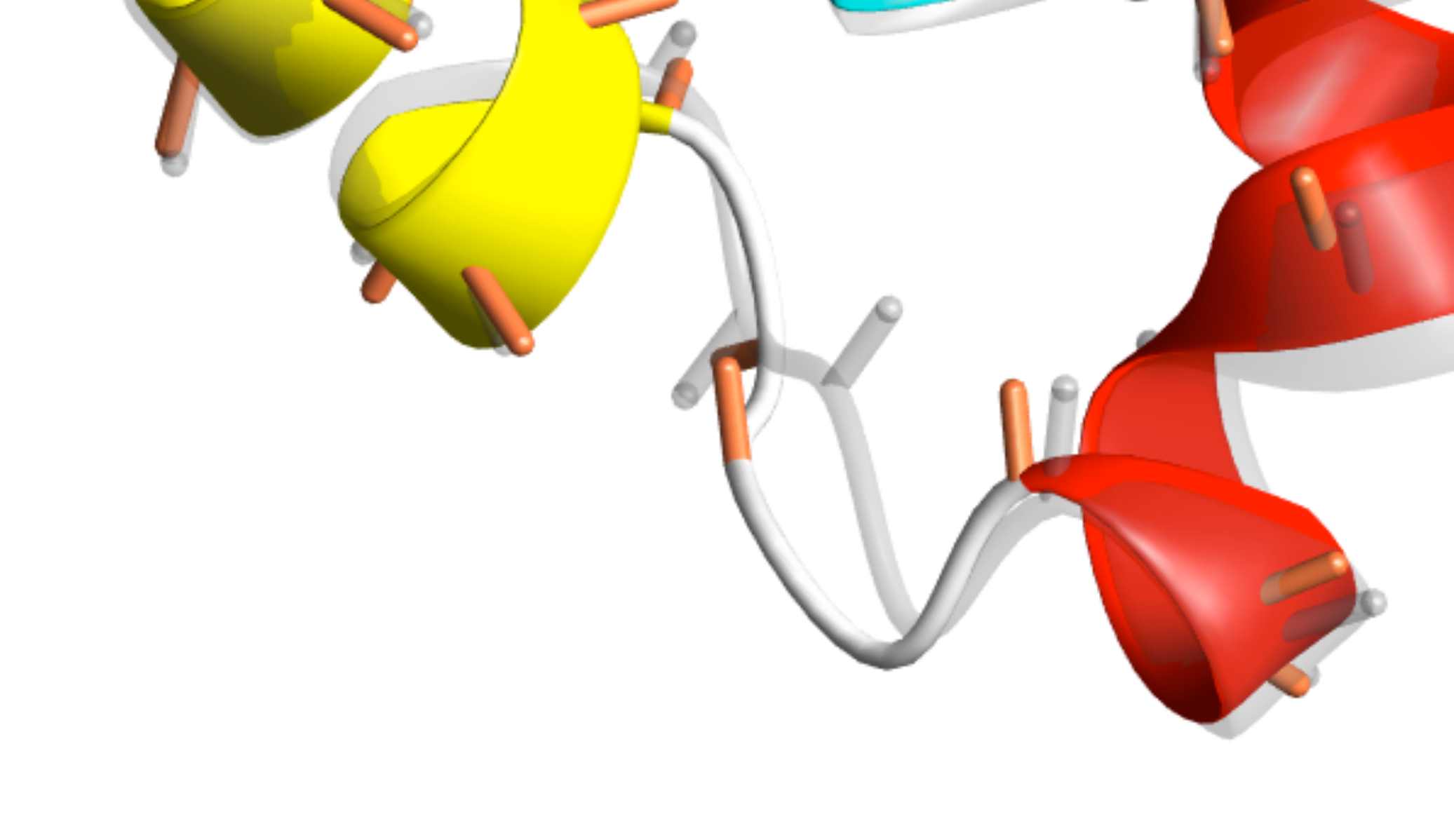}}
    \caption{(Color online)  Close-up of Figure 11 around site 31, the location of the second soliton. Multi-soliton is opaque,
    the PDB structure of 1LMB is translucent.}
    \label{fig:simple}
  \end{center}
\end{figure}

Finally, in Table III we list the parameters in (\ref{E0})-(\ref{ED}), for all the seven solitons.
\begin{table}[h]
\begin{center}
\caption{Our best parameter values for the multi-soliton solution  that models 1LMB. Notice that in line with
(\ref{malpha}), (\ref{mbeta}), the values of $m_1$, $m_2$ imply that  all 
the regular structures are $\alpha$-helices except for the one separating 
solitons 4 and 5 which is a short $\beta$-strand. We also note that the overall scale of the parameters
is fixed by the normalization of the first term in (\ref{EA}) which we have chosen for convenience.
}
\vspace{3mm}
\begin{tabular}{c|cccc}
\\ [-3mm]
Soliton & $q_1$ & $q_2$ & $m_1$ & $m_2$  \\
\hline \\ [-3mm]

1 & ~ 1.12091 ~ & ~ 1.87372  ~ & ~ 1.55737 ~ & ~ 1.50013 ~   \\
\hline \\ [-3mm]

2 & 0.357906 & 9.69166 & 1.65666 & 1.54119  \\
\hline \\ [-3mm]

3 & 0.260909 & 6.14144 & 1.68182 & 1.54676  \\
\hline \\ [-3mm]

4 & 0.684119 & 4.75578 & 1.47243 & 1.09234 \\
\hline  \\ [-3mm]

5 & 5.15882 & 6.77828 & 1.07066 & 1.53464  \\
\hline \\ [-3mm]

6 & 0.314503 & 0.38534 & 1.66164 & 1.6235  \\
\hline \\ [-3mm]
7 & 0.947322 & 0.624884 & 1.58568 & 1.51678  \\
\hline 

\end{tabular}

\vskip 0.4cm

\begin{tabular}{c|cccc}
\\ [-3mm]
Soliton & $a$ & $b$ & $c/2$ & $d/2$  \\
\hline \\ [-3mm]

1 & ~ -1.0     ~ & ~ -32357500 ~ & ~ 47.5340 ~ & ~ 438750 ~   \\
\hline \\ [-3mm]

2 & -1.0    & -1.86689   & 0.0413976   & 0.00101952  \\
\hline \\ [-3mm]

3 & -1.0     & -9.26604   & 0.121919  & 0.0271569  \\
\hline \\ [-3mm]

4 & -1.0   & -7.51371  & 0.013751  & 0.0148377  \\
\hline \\ [-3mm]

5 & -1.0   & -25.0125  & 0.251803  & 0.597751  \\
\hline \\ [-3mm]

6 & -1.0   & -23.9299   & 0.0181312 & 0.0410994  \\
\hline \\ [-3mm]

7 & -1.0    & 7.9809   & 0.000215793  & 0.037091 
 \\ 

\hline 
\end{tabular}

\vskip 0.4cm

\begin{tabular}{c|cccc}
\\ [-3mm]
Soliton & $a_\theta $ & $ b_\theta \cdot 10^{-12}$ & $c_\theta $ & $d_\theta $  \\ 
\hline \\ [-3mm]
1 & ~ 1.24035  ~ & ~ -475.728  ~ & ~ 1.0 ~ & ~ 5.44473$\cdot 10^{-10}$   ~   \\ [1mm]
\hline \\ [-3mm]

2 & ~ 2.74724    ~ & ~ -198463  ~ & ~ 1.0 ~ & ~ 0.42667  ~   \\ [1mm] 
\hline \\ [-3mm]

3 & 1.32293    & -5.021439$\cdot 10^{8}$   & 1.0 & 3.74966$\cdot 10^{-6}$   \\ [1mm]
\hline \\ [-3mm]

4 & 1.34998   & -81.1641  & 1.0  & 6.59854$\cdot 10^{-9}$  \\ [1mm]
\hline  \\ [-3mm]

5 & 1.38447  & -6629519   & 1.0 & 5.20625$\cdot 10^{-7}$  \\ [1mm]
\hline \\ [-3mm]

6 & 1.38293   & -6.28532  & 1.0 & 1.34992$\cdot 10^{-5}$ \\ [1mm]
\hline \\ [-3mm]

7 & 1.23869    & -177.296    & 1.0  & 7.6271$\cdot 10^{-7}$
 \\ 

\hline 
\end{tabular}

\vskip 0.4cm

\begin{tabular}{c|cccc}
\\ [-3mm]
Soliton & $a_\varphi$ & $ b_\varphi$ & $c_\varphi $ & $d_\varphi \cdot 10^{-10}$  \\ 
\hline \\ [-3mm]
1 & ~ 0.971374  ~ & ~ -6.7206$\cdot 10^{-10}$  ~ & ~ 1.0 ~ & ~ 7211.16   ~   \\ [1mm]
\hline \\ [-3mm]

2 & ~ 0.813254   ~ & ~ 2.3582$\cdot 10^{-7}$ ~ & ~ 1.0 ~ & ~ 1481.99 ~   \\ [1mm] 
\hline \\ [-3mm]

3 & 0.771272   & 2.74 $\cdot 10^{-6}$   & 1.0 & -51.6611  \\ [1mm]
\hline \\ [-3mm]

4 & 0.616865   & 1.2158$\cdot 10^{-11}$  & 1.0  & 5478  \\ [1mm]
\hline  \\ [-3mm]

5 & 0.89315  & 3.87$\cdot 10^{-9}$   & 1.0 & 1039.56  \\ [1mm]
\hline \\ [-3mm]

6 & 0.545154  & 0.184472  & 1.0 & 1.2685 \\ [1mm]
\hline \\ [-3mm]

7 & 0.988183    & 3.12459$\cdot 10^{-9}$   & 1.0  & 2.09412
 \\ 

\hline 
\end{tabular}

\end{center}
\label{solenoid}
\end{table}

%
%
%
%
%
%
%
%
%
%
%
%
%
%
%

\section{IV: Collapse studies of 1LMB}

In the backbone  energy  (\ref{EA}), (\ref{EB}) we have retained only 
those  variables that are relevant to our 
description of the C$_\alpha$ geometry. 
The derivation of (\ref{EA}), (\ref{EB}) is based on the general
concept of universality \cite{widom}-\cite{fisher}, in combination with the
requirement that the energy  must be independent of  the coordinate frame
where it is computed. Consequently, {\it by construction},  
our energy function  correctly describes
the leading order long distance contribution to {\it any} energy function that is 
grounded on more detailed atomic level considerations.   
{\it All} the variables and interactions that are less relevant for the
description of the C$_\alpha$ geometry, are accounted for 
effectively through the functional form and the 
parameter values  of the individual contributions to (\ref{EA}), (\ref{EB}).  

We shall try and approach protein dynamics in the same universal
manner. We average over all very short time scale 
oscillations, vibrations and other 
tiny fluctuations in the positions individual atoms that are basically irrelevant to the 
way how the folding progresses over those time scales that are 
biologically relevant. The general concept of universality \cite{widom}-\cite{fisher}
proposes us to adopt a  Markovian Monte Carlo time evolution 
with the following universal, coarse grained  heat bath  probability 
distribution \cite{glauber}, \cite{lebo}, \cite{marty}
\begin{equation}
\mathcal P = \frac{x}{1+x} \ \  \ \ {\rm with}  \   \ \ \ x =     \exp\{ - \frac{ \Delta E}{kT} \}  
\label{P}
\end{equation}
Here $\Delta E$ is the energy difference between consecutive MC time steps,  that we 
compute from (\ref{EA}), (\ref{EB}). We select the numerical value of the temperature factor 
$kT$ so that the model describes the appropriate phase. In \cite{max} it has been shown 
that (\ref{EA}), (\ref{EB}) is capable of describing the three
phases of polymers. At low values of $kT$  we are in the phase of collapsed proteins.
As the value of $kT$ increases and reaches the  $\Theta$-point value, there is a
transition to random coil phase. When the temperature
reaches even  higher values, there is a cross-over to self-avoiding random walk. 

It turns out that in the collapsed phase, 
below the $\Theta$-point temperature, the universal aspects of folding dynamics
are quite independent of the numerical value of $kT$. For concreteness,
we perform our simulations in the collapsed using the value
 \[
kT  = 10^{-15}
\]
Note that the overall normalization of $kT$
can always be changed by an overall normalization of the parameters in Table III. 

We shall assume that during the folding process 
there are no re-arrangements  in  the backbone covalent bond 
structure, such as chain crossings. For this we 
introduce a self-avoidance  condition that keeps the distance 
between any two backbone C$_\alpha$ atoms 
at least as large as the length of a typical van der Waals radius which is around
$\sim$1.3 \AA ngstr\"om. 

Note that we do not propose that 
(\ref{P}) is capable of describing the atomic level dynamics of the folding process.
Such minuscule details  are highly sensitive to the initial atomic configuration. 
A detailed knowledge of the time evolution of a
particular atom during the collapse can hardly have any practical relevance for
the underlying physical principles and phenomena.  Thus, 
for the purpose of conceptually understanding the temporal evolution of a protein towards its 
native conformation, the dynamics described by  (\ref{P}) is sufficient. We argue that the combination 
of (\ref{E0}),  (\ref{P}) correctly captures  the universal statistical aspects of protein collapse 
over the biologically relevant temporal and spatial scales.

%
%
%
%
%
%
%
%
%
%
%
%
%
%
%

\subsection{A: Antiferromagnetism and folding nuclei }

In our approach, proteins have a modular structure. A folded protein is built by combining together solitons
of the discrete non-linear Schr\"odinger equation (\ref{nlse}), one after another.   
From the point of view of the energy function (\ref{EA}), (\ref{EB}), a uniform helical configuration 
is one with
\[
\kappa_i \, \approx \, m
\]
and with the  value 
of  $\tau_i$ computed from (\ref{tauk}) this is a ground state of the energy. In particular, 
a straight linear rod is a special case, it is a ground state when $m =0$. 

As usual, the physical principles that
give rise  to protein folding are best analyzed in the absence of other processes and interfering 
agents. For this reason, in the present sub-section,  we study the soliton formation along 
a helical backbone, by inspecting 
how the folded configuration builds from 
the ground state of the energy. 
Consequently we use a straight helical structure as our initial configuration. In the case of the 
$\lambda$-repressor protein we are particularly interested in the formation of the third, DNA binding soliton.

The parameter values in Table III are uniform over each putative super-secondary structure.
In particular, there is no information on the loop locations.  In a protein, the placement of a
loop often correlates with the position of certain 
amino acids, such as proline and glycine, that act as folding nuclei.
In order to model a folding nucleus we introduce a transient parameter 
$\sigma$ that sends the first term in (\ref{EA})
into
\begin{equation}
- \sum\limits_{i=1}^{N-1}  2\, \kappa_{i+1} \kappa_{i} \ \to \  \sum\limits_{i=1}^{N-1}  (2 \sigma-1) \cdot 2 
\, \kappa_{i+1} \kappa_{i}
\label{af}
\end{equation}
Initially, we set $\sigma = 0$ for all $i$ except at the links ($i,i+1$) along which the 
putative soliton centers are located. At these links we start with $\sigma=+1$. 
This corresponds to an anti-ferromagnetic {\it i.e.} repulsive nearest neighbor interaction
between the ensuing two sites. During the early stages of the simulation, 
we decrease the values of $\sigma$  so that after some number of steps we 
reach the uniform final value
$\sigma = 0$ for all links along the entire chain. 
 
For each super-secondary structure the value of $m$ in Table III determines  
the regular secondary structure which is located
either before or after the corresponding  loop, and the value 
of the parameter $q$ in (\ref{EA}) determines the propensity of this structure to form. 
The stability of $\alpha$ helices and $\beta$ strands is  
due to hydrogen bonds that form during the collapse, and consequently the value of $q$ can be interpreted
as a measure of
the strength of hydrogen bond interactions. In our simulations we wish to start from an initial configuration
with no initial hydrogen bonds. We conform to this  
by setting all $q = 0$ initially.  
We then switch on the hydrogen bond interactions  by  increasing the values of the parameters $q$  
to those given in Table III.  We find that this stabilizes the regular secondary structures.

We wish to investigate the effects that the temporal ordering of loop formation  has on
folding and on misfolding. For this we 
compare different orderings in removing $\sigma $ and in 
switching on the values of the $q$.

 In the case of the lysogeny maintaining $\lambda$-repressor protein
we have considered various scenarios to conclude that there is the following general pattern: 
The seven solitons that we display in Figure 4 (right column)  
tend to form as pairs, with (2,3), (4,5) and (6,7) each a soliton-soliton pair.  
The first soliton is also made with a pair. But after the formation,  the pair of this
soliton moves away and disappears through the $N$-terminal
of the backbone. The alternative, where the first and second soliton form as a pair seems to give
rise to a misfolded state that furthermore appears to be unstructured.
We now describe in detail two generic examples that illustrate  this general pattern:
\vskip 0.2cm

{\it First example:} In our first generic example we start from a 
uniform, straight helical structure. In the Figure 13a we show the initial
$\kappa_i$ profile.
We note that  there is substantial 
latitude in choosing the initial values of $\kappa_i$ and $\tau_i$.
To begin with, we also set all $q=0$ so that there are no hydrogen
bonds to stabilize the helical structure. All the remaining parameters have the values that are 
listed in Table III. 
We introduce an antiferromagnetic coupling $\sigma=+1$ at links ($i,i+1)$ with 
$i=16, \, 23, \,33, \, 46, \, 49,\, 63,\, 67$. This models the folding nuclei at
the putative positions of the centers of the solitons. 
During the first part of the simulation, say during the first $2,000,000$ Monte Carlo steps, 
we adiabatically 
remove the folding nuclei by linearly decreasing the values of $\sigma$ until we reach
the final ferromagnetic values $\sigma=0$ at all sites. 
At the same time we turn on the hydrogen bonds, by increasing the values of the couplings
$q$ from zero to the values given in Table III. After  the first $2,000,000$ steps
all parameters along the backbone then
have the values shown  in Table III. 

We remind that according to our observations there is a lot of latitude in details of 
the procedure. 
\begin{figure}[!hbtp]
  \begin{center}
    \resizebox{8.cm}{!}{\includegraphics[]{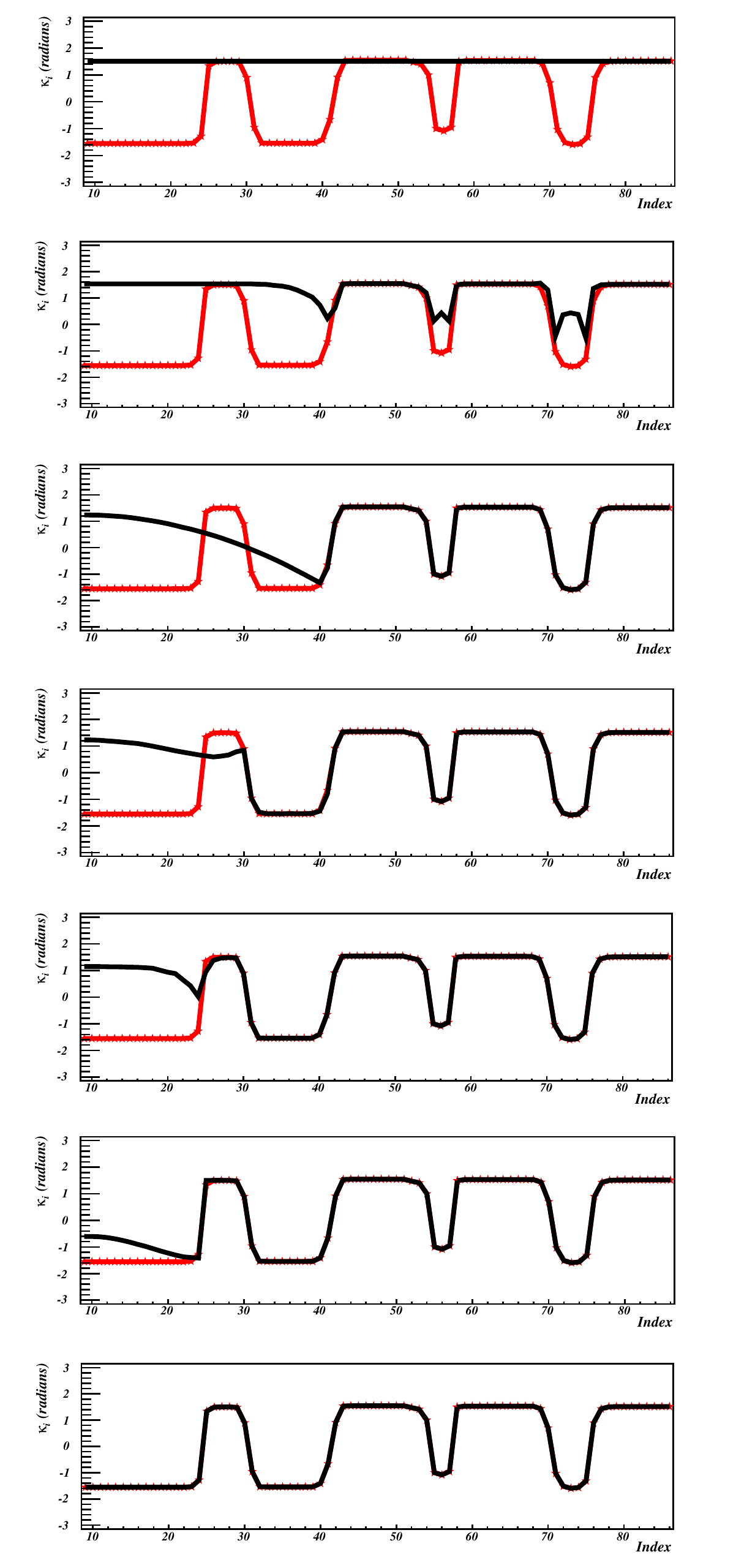}}
    \caption{(Color online) Time series of soliton (loop) formation along the 1LMB backbone in our simulation. The black line
    shows the time evolution of the solitons, the grey (red) line is the PDB profile. Time increases
    from top down. In the first Figure from the top we have the initial configuration, a straight helical structure.
    The solitons 4-7 form as the two soliton pairs (4,5) and (6,7). 
    At the position of soliton 3, there is also a pair formation. But
    the pair of soliton 3 propagates  along the chain towards the $N$-terminal,  until it 
    becomes anchored at site $i=23$ where it forms 
    the second soliton. A new soliton pair forms at site 16 and one of these two solitons 
    moves and disappears through the $N$-terminal,  leaving us with 
    the first soliton and the correctly folded backbone}
    \label{fig:simple}
  \end{center}
\end{figure}

We find that the last four solitons form as the pairs (4,5) and (6,7). At the putative 
location of the third soliton, we also observe the formation of a soliton-soliton pair. 
But the soliton which is located closer to the $N$-terminal
moves towards left as shown in Figure 13,  and becomes anchored at the site 23 where it forms
the second soliton of 1LMB. A new soliton pair appears at the putative location 
of the first soliton,  one of these solitons disappears through the $N$-terminal 
and the entire backbone stabilizes rapidly into the  correct native state. See Figure 13. 

\vskip 0.2cm

{\it Second example:}  In this example we simulate a scenario where the first pair is formed
before the third soliton. The initial configuration is a folded structure, with the fully formed soliton
pairs (1,2), (4,5) and (6,7)  {\it i.e.} these solitons have the same ($\kappa_i, \tau_i$) 
values as the 1LMB. But the DNA binding third soliton is absent and instead there is a 
helix extending from the second to the 
fourth soliton, see Figure 14 and 15. 
%
%
%
\begin{figure}[!hbtp]
  \begin{center}
    \resizebox{8.cm}{!}{\includegraphics[]{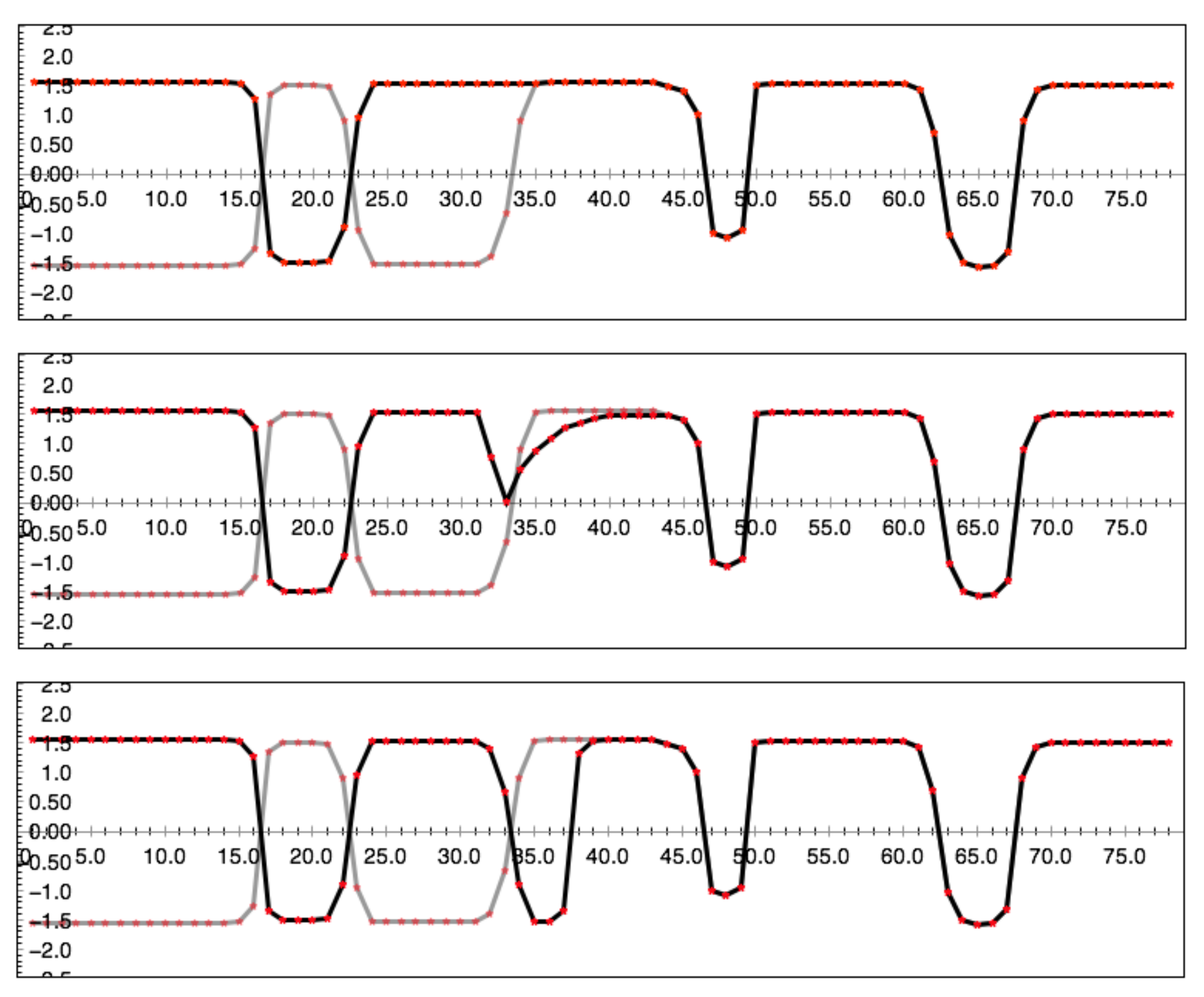}}
    \caption{(Color online)  If hydrogen bonds form slowly, there is an extraneous  soliton that disturbs
    binding between 1LMB and DNA, and probably no binding is possible. Grey line is the 1LMB profile, and black 
    with red dots is  the simulated profile.
    }
    \label{fig:simple}
  \end{center}
\end{figure}
%
%
%
\begin{figure}[!hbtp]
  \begin{center}
    \resizebox{8.cm}{!}{\includegraphics[]{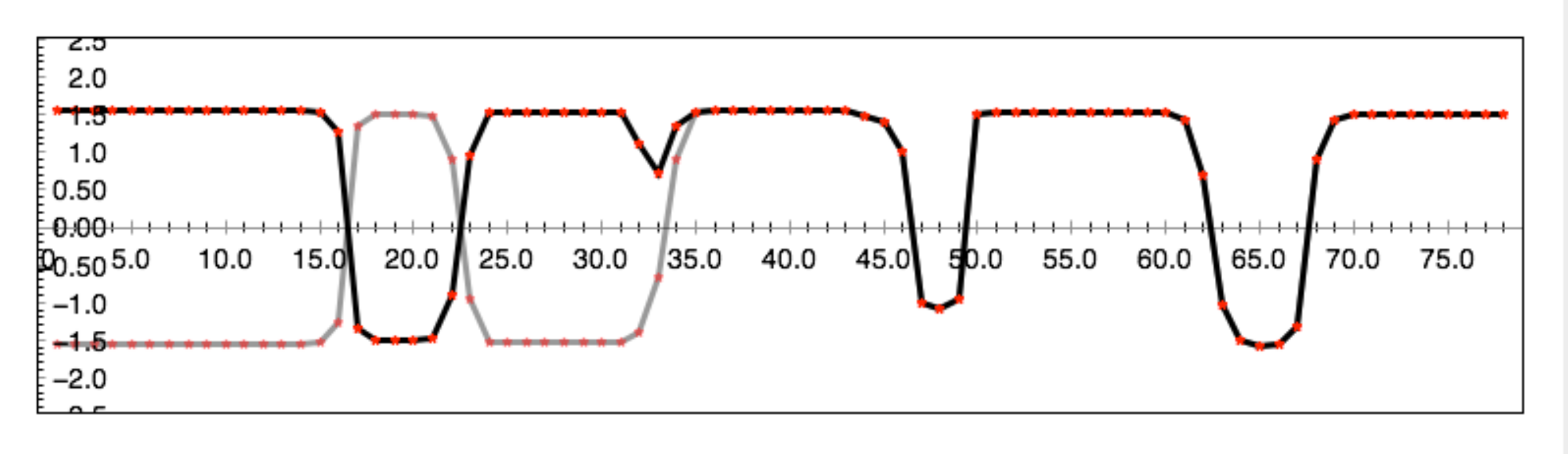}}
    \caption{(Color online)  If the hydrogen bonds are formed early, the backbone enters in an unstructured
    state where the 3 soliton become mis-formed. Consequently there is no soliton that would dock the 1LMB with DNA.
     Grey line is the 1LMB profile, and black 
    with red dots is  the simulated profile.
}
    \label{fig:simple}
  \end{center}
\end{figure}
The simulation starts with an antiferromagnetic $\sigma=1$
at link $i=34$, and with no hydrogen bond interactions {\it i.e.}
$q=0$ between second and fourth solitons. There is then an initial production
of a soliton-soliton pair as seen in Figure 15b and we find two possibilities:

If the hydrogen bonds form slowly
{\it i.e.} $q$ grows to its final value slowly in comparison to the removal of $\sigma$,  we arrive at a final
fold where there is an additional soliton (loop)
around site 39. We display the bond angle profile in Figure 14c.  The protein is
now misfolded. On the other hand, if the hydrogen bonds are formed 
more rapidly, we arrive at a configuration
where the third and fourth soliton annihilate each other. No soliton is formed, but 
due to steric restraints there is a slightly irregular helical region between sites 31 and 45. The
final configuration is shown in Figure 15.

We conclude that  if the ($1,2$) pair forms before the third soliton, 
the protein becomes misfolded into a state with more than one conformational substrate. We
propose that  this could be confirmed experimentally. It might relate to
the transition from the lysogenic to the lytic state in $\lambda$-phage. 

%
%
%
%
%
%
%
%
%
%
%
%
%
%
%
\subsection{B: Comparison of $\lambda$-repressor and CRO soliton structures}

In Figure 16 we compare the first two solitons in 1LMB to the first loop in the CRO  regulator protein that 
controls the transition to the lytic state. For the latter, we use the PDB structure with code 
2OVG.   In the CRO protein, the first loop is topologically more stable, 
in the sense that it consists of a single soliton. As such the loop in 2OVG is more stable than in 1LMB.
For a plane curve, a single soliton can be made or deleted  only by transporting 
it through one of the end points of the curve. On the other hand,  a pair of solitons  such as the one in the left hand side of Figure 16 is not topologically stable but can be more easily created or removed
locally,  by a saddle-node bifurcation that brings the two solitons  together. This  
removes the corresponding loop by converting it (in this case)  into a single long $\alpha$-helix. 

A comparison between the $\lambda$-repressor and CRO  profiles  in Figure 16 
proposes the following  mechanism for the lysogenic-lytic transition: 
Under lysogenic conditions where the $\lambda$-repressor 
protein prevails, the soliton pairs of the $\lambda$-repressor protein that are located 
immediately prior and after the DNA binding domain are relatively stable. But when there is a change in the 
environmental conditions that excites phonon fluctuations along the protein chain such as raise in temperature or UV radiation, or maybe an enzymatic action that remains to be identified, either of these  soliton pairs can 
discharge by a saddle-node bifurcation.
This  bifurcation disturbs the structure of the immediately adjacent DNA binding motif  to the extent that 
the protein looses its capability to maintain the lysogenic phase. 
Since each of the corresponding motifs in the CRO protein are  topologically more stable single soliton
configurations, they are much more insensitive
to effects such as local phonon excitations due to UV radiation and thermal effects,  and consequently the lytic phase can take over. 
\begin{figure}[!hbtp]
  \begin{center}
    \resizebox{8.cm}{!}{\includegraphics[]{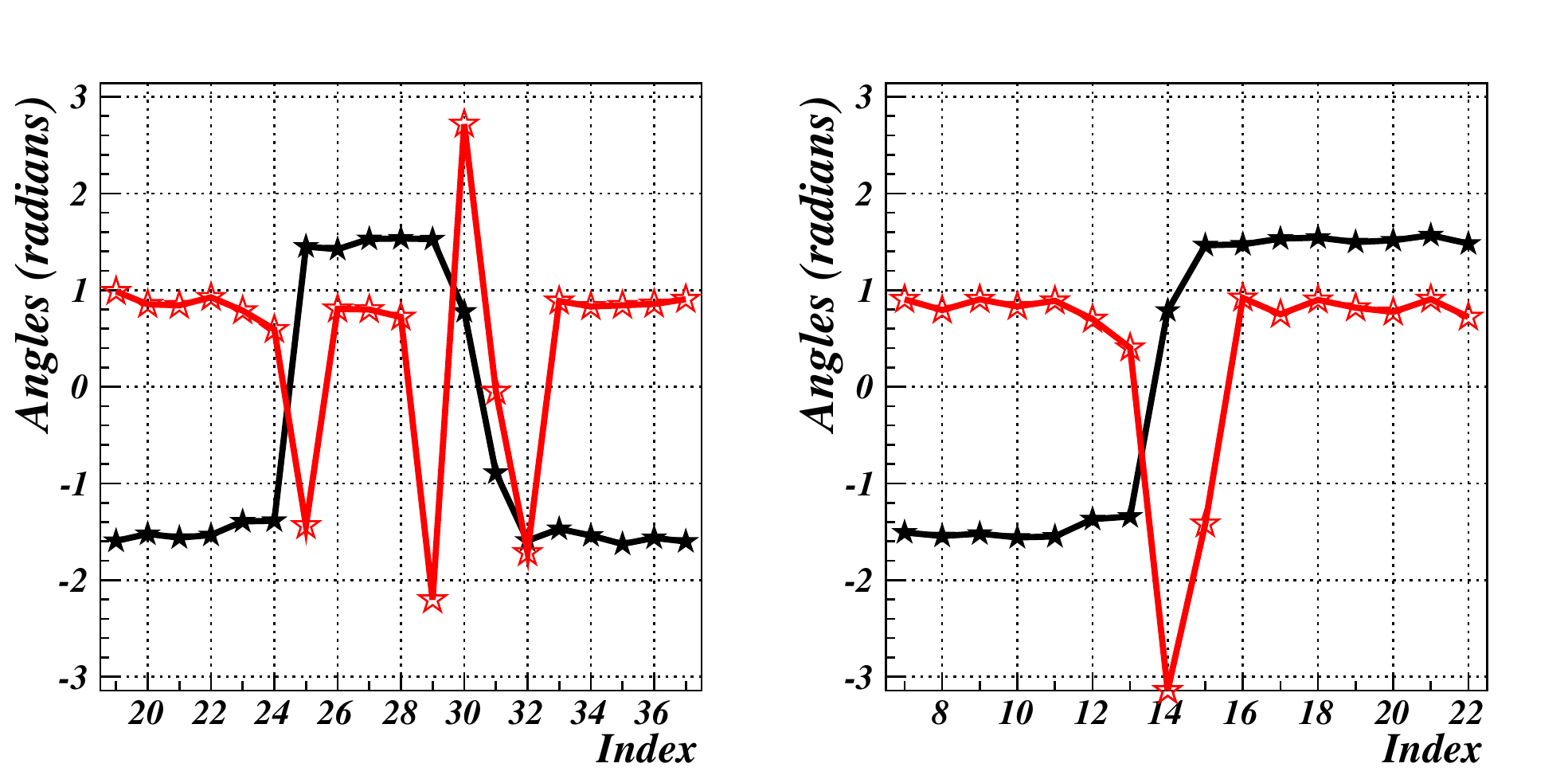}}
    \caption{(Color online)  The resolved ($\kappa_i, \tau_i$) spectrum for the first PDB helix-loop-helix of 1LMB (left) and the corresponding structure of in 2OVG (right). 
    The bond angle  $\kappa$ is black, torsion angle  $\tau$ is grey (red). The bond angle
    spectra reveal that in 1LMB the loop is a bound state of two close-by
    solitons    while in 2OVG there is only one soliton profile.}
    \label{fig:simple}
  \end{center}
\end{figure}
we display the 
first PDB helix-loop-helix motif and for comparison we display the corresponding structure  in  the CRO 
protein with PDB code 2OVG.  In the case of  $\lambda$-repressor the motif
is clearly a bound state of two solitons while in the case of  CRO we have a single isolated soliton.

%
%
%
%
%
%
%
%
%
%
%
%
%
%
%

\subsection{C: Heating and cooling in 1LMB }

We apply (\ref{P}) to theoretically investigate what takes place in 1LMB when 
we heat it into the random coil phase, and then re-cool it back to the collapsed
phase. According to Anfinsen \cite{anf} the protein should  return
to its original conformation.

We start from the multi-soliton configuration that describes the PDB structure 1LMB with
parameter values given  in Table III.  Unlike in the previous subsection, 
during the entire heating and cooling cycle
we now keep all the parameter values intact. In particular, neither during
the heating nor during the cooling do we introduce any transient 
antiferromagnetic parameter $\sigma$  as in the previous subsection. 
Nor do we change the parameter values $q$ during the present 
process.  As a consequence the position of any soliton and the size of any helical structure
becomes determined dynamically, without any explicit folding nuclei. Moreover,
since the parameter values 
$q$ do not change, the strength of the underlying  hydrogen bond interactions 
remains constant during the simulations. Any deformation or adjustment in the helical structures 
will be entirely due to thermal fluctuations during the heating and cooling cycle.

We introduce the heat bath dynamics (\ref{P}). 
The temperature $kT$ is assumed to be globally determined, and 
the heating and cooling should proceed slowly over the  
biologically relevant time scales.  This ensures that the entire 
protein structure is kept at an equal temperature value so that we can
ignore any effects due to local temperature variations.

During heating and cooling cycle we follow the backbone evolution by computing the 
root-mean-square distance (RMSD) between the C$_\alpha$ coordinates $\mathbf r_{0i}$ of
the 1LMB backbone and those of the multi-soliton configuration $\mathbf r_i$, as a function
of the temperature
\begin{equation}
RMSD(T) \ \buildrel {def} \over {=} \ \sqrt{ \frac{1}{N} \sum_i ( \mathbf r_{i0} - \mathbf r_i)^2 }
\label{rmsd}
\end{equation}
We start from a low temperature value that corresponds to the collapsed phase. We again
choose the numerical value
\begin{equation}
kT \, = \, 10^{-15}
\label{lowT}
\end{equation}
for the initial configuration. We adiabatically increase the temperature $kT$ to some high value 
during 500,000 MC steps. We then keep the system 
at this high temperature during 1,000,000 MC steps, and finally cool it back to the original
temperature value (\ref{lowT}) during another 500.000 steps.   
The  relatively large number of MC steps in our cycle  
at the high temperature ensures that the protein  becomes fully thermalized to that temperature value.
 
We have made simulations with several hundreds of repeated heating and cooling cycles, always
starting with (\ref{lowT}) and heating to different high temperature values that are 
well above the $\Theta$-point, where the protein enters the random walk phase.  
From Figure 17 we learn
%
%
%
\begin{figure}[!hbtp]
  \begin{center}
    \resizebox{8.cm}{!}{\includegraphics[]{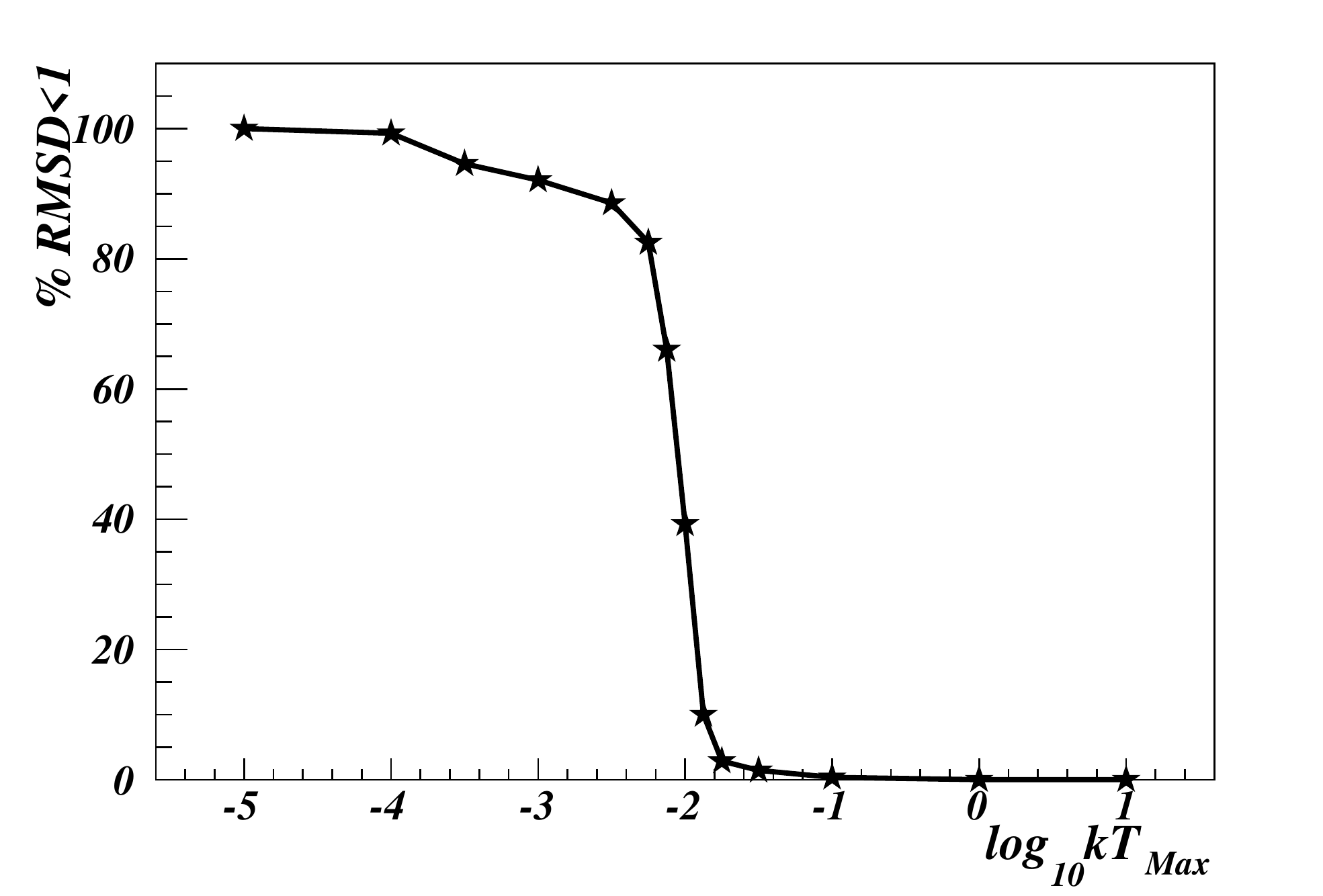}}
    \caption{At temperature values $\log kT <  -4$ the protein returns to its original conformation
    after the heating and cooling cycle. Between $-4 < \log kT <  -2$ a small fraction starts to misfiled.
    At $\log kT \approx -2$ there is a rapid transition, so that proteins that have been heated to higher
    temperatures become misfolded.
    }
    \label{fig:simple}
  \end{center}
\end{figure}
that as long as the high temperature
value remains below 
\begin{equation}
kT_{C1} \, \approx \, 10^{-4}
\label{C1}
\end{equation}
the protein always returns back to its original shape upon cooling.  But between values in the range
\[
10^{-4} \approx kT_{C1} < kT < kT_{C2} \approx 10^{-2} 
\]
there is a small fraction of cycles at the end of which the protein becomes misfolded. Finally at around
the temperature value $kT_{C2}$ there is a rapid cross-over and if the heating temperature
exceeds 
\[
T > T_{C2}
\]
the final conformation is  always misfolded. The misfolding is caused by deformation of one or more of the loops, 
often due to the wrong ordering in loop formation.

In Figure 18 we show as an example, how (\ref{rmsd}) evolves in average 
during a typical heating cycle that reaches very near the critical
temperature value (\ref{C1}). We find that during the heating there is a rapid increase  in the 
RMSD distance. This  is due to the transition to the random coil phase. When the system
is kept at the high temperature, there are substantial thermal fluctuations. The shaded region (blue online)
denotes the
one standard deviation extent of these fluctuations. During the cooling we have
a collapse transition, and at the end the value of (\ref{rmsd}) becomes small with practically no fluctuations,
indicating that the protein has 
returned to the original 1LMB like conformation.  In Figure 19 we display a generic random coil structure
that we observe during the heating phase. It has the look of a typical random coil with no regular,
helical structure remaining. The structure is not static, but subject to very strong thermal fluctuations in its shape.
%
%
%
\begin{figure}[!hbtp]
  \begin{center}
    \resizebox{8.cm}{!}{\includegraphics[]{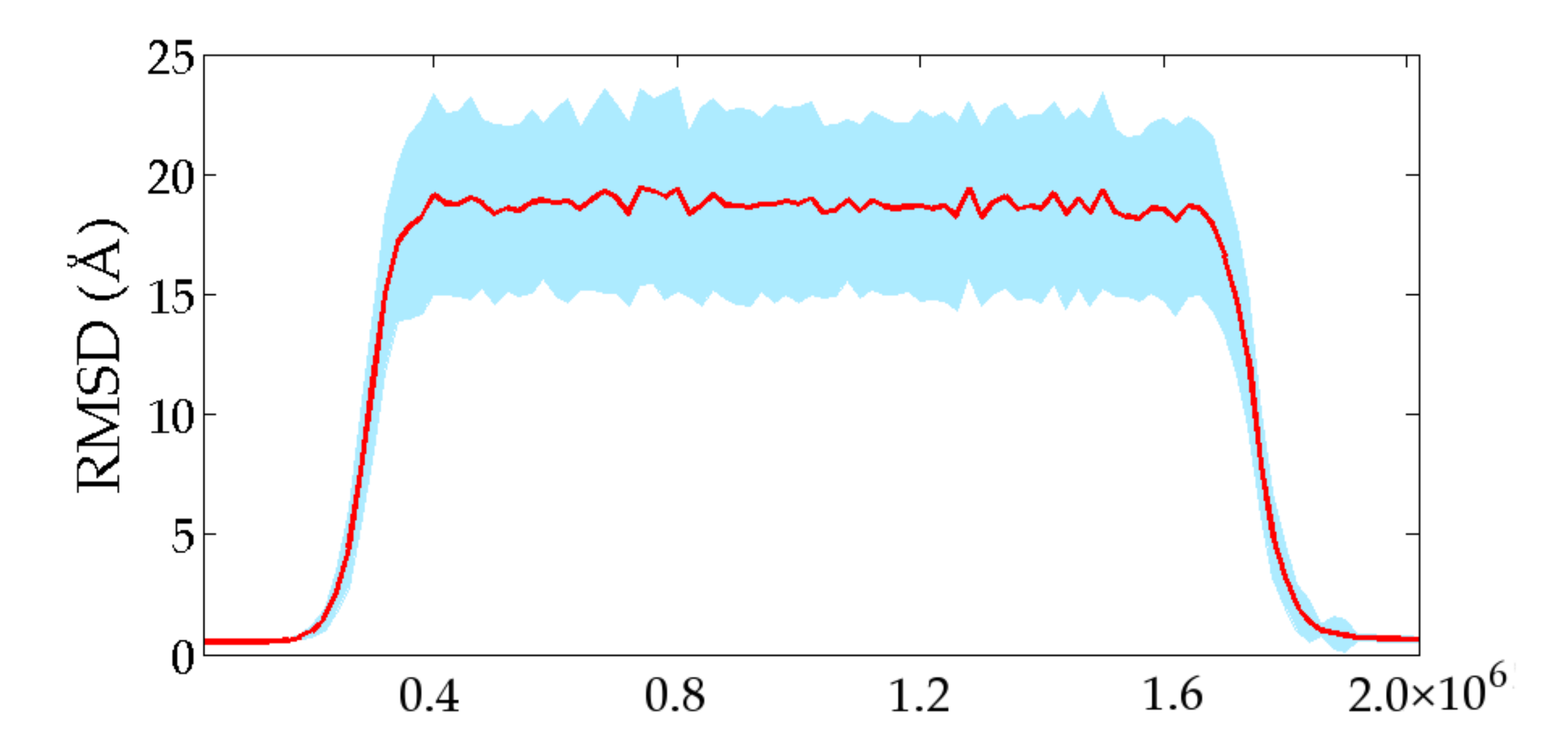}}
    \caption{(Color online)  The evolution of the RMSD value (\ref{rmsd}) between the 1LMB and the heated structure during 
    one cycle. At the high temperature value which is here $kT = 10^{-4}$, the RMSD is large and subject
    to large  thermal fluctuations; the shaded area denotes the 
    one standard deviation fluctuation regime in the values of (\ref{rmsd}). At the end of the cycle, 
    the protein folds back to the conformation of 1LMB. }
    \label{fig:simple}
  \end{center}
\end{figure}
\begin{figure}[!hbtp]
  \begin{center}
    \resizebox{8.cm}{!}{\includegraphics[]{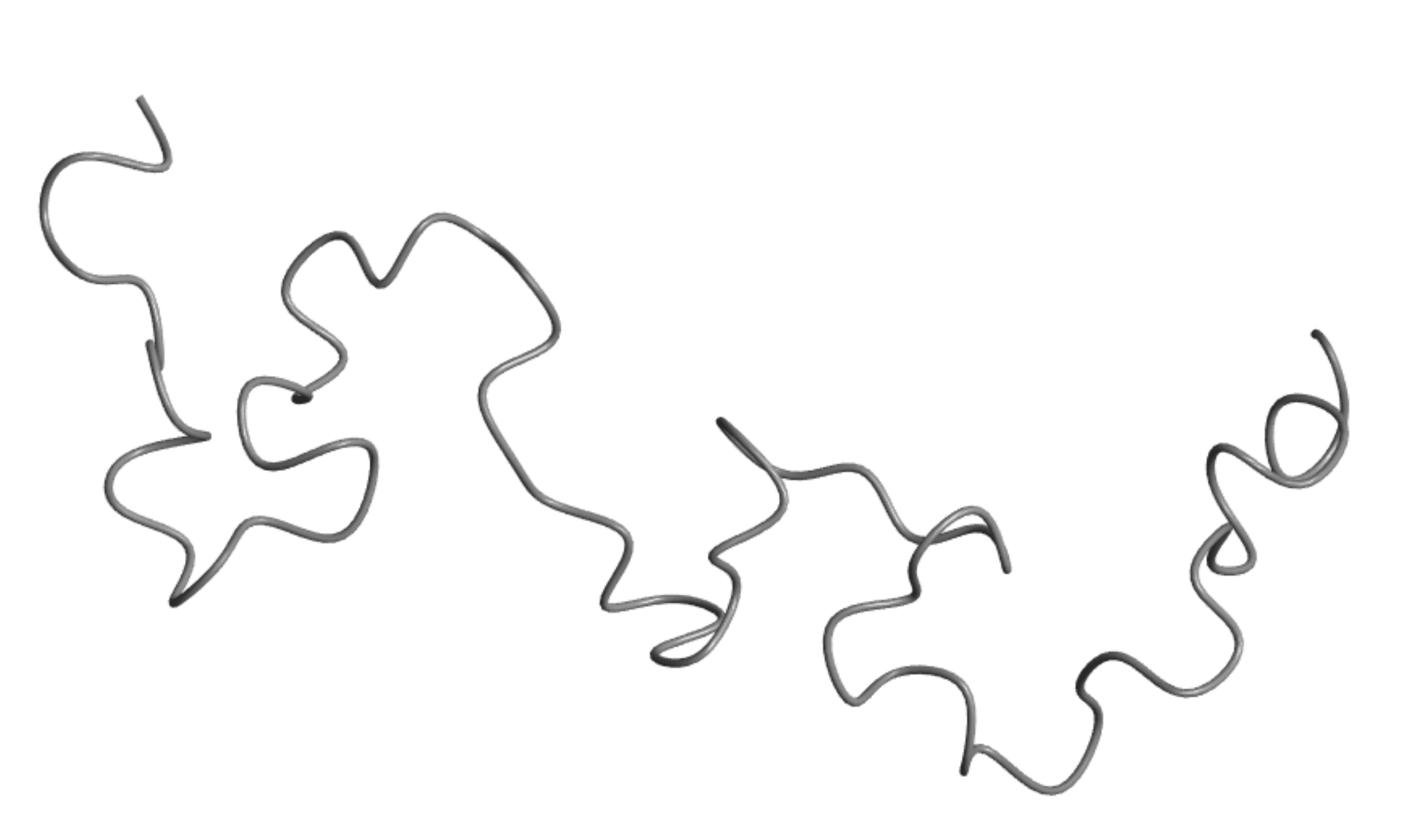}}
    \caption{A generic configuration during the high temperature 
    regime in the cycle of Figure 18. The structure  is an apparent random coil with no regular structure, and 
    its detailed form is subject to strong thermal fluctuations. }
    \label{fig:simple}
  \end{center}
\end{figure}

%
%
%
%
%
%
%
%
%
%
%
%
%
%
%

\section{Summary }

In summary, we have investigated the structure and folding pathways of the lysogeny maintaining
$\lambda$-repressor protein.
Our approach is based on an effective energy function, that we have argued describes the small
fluctuation limit of {\it any} atomic level energy function. Our justification of  the energy function is entirely based
on two very general physical principles: The concept of universality originally introduced in the context of phase transitions
and critical phenomena, and the demand that any physical quantity must be independent of the coordinate system
that is used for its description. Using these two universal physical principles, we have shown that the long wavelength 
fluctuation limit of the energy function for both the backbone C$_\alpha$ and side-chain C$_\beta$ atoms is fully 
determined in an essentially unique fashion. 

The energy function we have derived, computes the  C$_\alpha$ and C$_\beta$ conformation in terms of a
soliton solution to a variant of the discrete non-linear Schr\"odinger equation as the modular component. 
The explicit
form of the relevant dark soliton solution is not known to us, but we have found that an excellent approximation
can be obtained by naively discretizing the known dark soliton of the continuum non-linear Schr\"odinger equation. 
We are able to re-construct the entire backbone of the $\lambda$-repressor 
protein with a precision that compares and even exceeds 
the precision of the experimental crystallographic structure with PDB code 1LMB, when we determine the precision in terms of
the Debye-Waller B-factor fluctuation distance. The high precision of our theoretical description enables us to conclude
that the second soliton solution that appears in our description of the 1LMB is {\it unique} to this protein, there are
no other similar solitons in the entire Protein Data Bank. The remaining solitons including the DNA binding one are
all ubiquitous in PDB.  We have also investigated the corresponding soliton structure in the CRO protein that
is responsible for the lytic phase, and found that this soliton appears more stable and is also commonly found in
the PDB data. These observations suggest that the transition between the lysogenic and lytic life-cycles could 
somehow relate to the very exceptional structure of the second soliton in 1LMB.

We have extended our energy function to describe the collapse dynamics of 1LMB. In line with the construction of
the energy function, we rely on the concept of universality to propose that at biologically relevant time scales 
the folding dynamics can be described in terms of Glauberian relaxation dynamics, with Markovian time evolution. 
In this way we have found that
all solitons in 1LMB appear to form as soliton-soliton pair. In particular, the second soliton with its exceptional 
structure forms as a pair with the DNA binding third soliton; The pair of the first soliton flows away and disappears through
the $N$-terminal of the protein. Moreover, if for some reason the formation of the second and third solitons is disrupted,
for example if the second soliton forms by itself before the third soliton, we find that the third soliton can not form properly.
It is either formed in combination of an extra soliton, or then the protein enters in an unstructured conformation. The choice
between these two alternatives is made by the strength of the hydrogen bond formation.


\vskip 0.5cm
A.N.  thanks H. Frauenfelder and  G. Petsko  for communications and J. \AA qvist for discussions.


\end{document}